\documentclass[sigconf]{acmart}

\usepackage{tabularx}
\usepackage{lipsum} 

\usepackage{fancyvrb,fvextra}
\usepackage{multirow}
\usepackage{epigraph}
\usepackage{enumitem}

\AtBeginDocument{%
  \providecommand\BibTeX{{%
    \normalfont B\kern-0.5em{\scshape i\kern-0.25em b}\kern-0.8em\TeX}}}

\setcopyright{acmlicensed}
\copyrightyear{2025}
\acmYear{2025}
\setcopyright{cc}
\setcctype{by}
\acmConference[CHI '25]{CHI Conference on Human Factors in Computing Systems}{April 26-May 1, 2025}{Yokohama, Japan}
\acmBooktitle{CHI Conference on Human Factors in Computing Systems (CHI '25), April 26-May 1, 2025, Yokohama, Japan}\acmDOI{10.1145/3706598.3713749}
\acmISBN{979-8-4007-1394-1/25/04}

\begin{document}

\title[Social-RAG]{Social-RAG: Retrieving from Group Interactions to Socially Ground AI Generation}

\author{Ruotong Wang}
\email{ruotongw@cs.washington.edu}
\affiliation{%
  \institution{University of Washington}
  \city{Seattle, Washington}
  \country{United States}}
\orcid{0000-0003-0964-6943}

\author{Xinyi Zhou}
\email{xzhou@cs.washington.edu}
\affiliation{%
  \institution{University of Washington}
  \city{Seattle, Washington}
  \country{United States}}
\orcid{}

\author{Lin Qiu}
\email{lq9@cs.washington.edu}
\affiliation{%
  \institution{University of Washington}
  \city{Seattle, Washington}
  \country{United States}}
\orcid{0009-0003-2354-1192}

\author{Joseph Chee Chang}
\email{josephc@allenai.org}
\affiliation{%
  \institution{Allen Institute of AI}
  \city{Seattle, Washington}
  \country{United States}}
\orcid{0000-0002-0798-4351}

\author{Jonathan Bragg}
\email{jbragg@allenai.org}
\affiliation{%
  \institution{Allen Institute of AI}
  \city{Seattle, Washington}
  \country{United States}}
\orcid{0000-0001-5460-9047}

\author{Amy X. Zhang}
\email{axz@cs.uw.edu}
\affiliation{%
  \institution{University of Washington}
  \city{Seattle, Washington}
  \country{United States}}
\orcid{0000-0001-9462-9835}
\renewcommand{\shortauthors}{Wang, et al.}

\begin{abstract}
 AI agents are increasingly tasked with making proactive suggestions in online spaces where groups collaborate, yet risk being unhelpful or even annoying if they fail to match group preferences or behave in socially inappropriate ways. 
Fortunately, group spaces have a rich history of prior interactions and affordances for social feedback that can support grounding an agent's generations to a group's interests and norms. 
We present Social-RAG, a workflow for socially grounding agents that retrieves context from prior group interactions, selects relevant social signals, and feeds them into a language model to generate messages in a socially aligned manner. 
We implement this in \textsc{PaperPing}, a system for posting paper recommendations in group chat, leveraging social signals determined from formative studies with 39 researchers.
From a three-month deployment in 18 channels reaching 500+ researchers, we observed PaperPing posted relevant messages in groups without disrupting their existing social practices, fostering group common ground.

\end{abstract}



\begin{teaserfigure}
\centering
  \includegraphics[width=.86\textwidth]{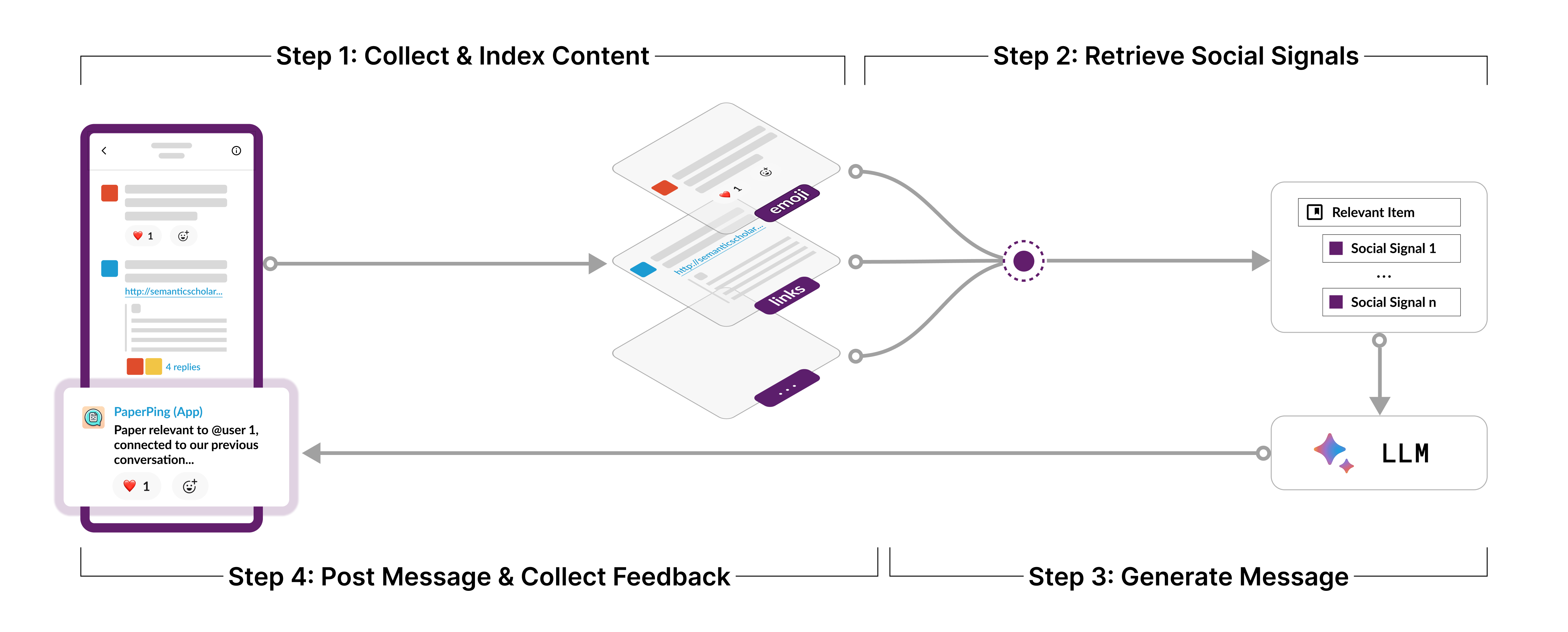}
  \caption{We introduce Social-RAG, a workflow for LLM-based AI agents to generate content aligned with group interests and norms. Step 1 involves collecting and indexing the group's conversation history into a social knowledge base. When tasked with making a post, relevant item suggestions (e.g., papers) and signals (e.g., group's topical interests) are retrieved and ranked (Step 2). Social-RAG then feeds these signals as context into an LLM to generate a socially grounded message (Step 3). The message is posted to group channels, where it can collect feedback that also gets indexed (Step 4).}
  \Description{We introduce Social-RAG, a workflow for LLM-based AI agents to generate content aligned with group interests and norms. The workflow contains four steps. After collecting and indexing group interaction history content into a social fact knowledge base (Step 1), Social-RAG analyzes and retrieves relevant item suggestions (e.g., papers) and signals (e.g., group's topical interests) from prior group interactions (Step 2). It then selects and ranks relevant social signals, which are fed as context into an LLM to generate a succinct message (Step 3). The message is then posted to social channels, where the agent can continue to collect and index group reactions (Step 4).}
  \label{fig:socialRAG}
\end{teaserfigure}



\def\authnote{1}
\newcommand{\fixme}[1]{\ifnum\authnote=1{\textcolor{red}{[FIXME: #1]}}\fi}

\newcommand{\xinyi}[1]{\textcolor{teal}{#1-Xinyi}}

\newcommand{\amy}[1]{\textcolor{orange}{AZ: #1}}

\newcommand{\jb}[1]{\textcolor{blue}{JB: #1}}

\newcommand{\jc}[1]{\textcolor{pink}{[JC: #1]}}

\newcommand{\rw}[1]{\textcolor{black}{#1}}

\maketitle

\section{Introduction}

AI-enabled agents are increasingly being  deployed in online social environments, including group conversational and collaborative spaces. They serve as assistants, facilitators, or collaborators to support varied tasks such as conversation summarization~\cite{zhangMakingSenseGroup2018a}, brainstorming~\cite{rayan2024exploring, shin2023introbot}, and conflict resolution~\cite{govers2024ai, tessler2024ai}. 
These agents often leverage AI to insert automated messages into group spaces where human-human conversation and interaction naturally occur; for instance, they can
post messages to facilitate collaborative learning in group chat channels~\cite{caiAdvancingKnowledgeTogether2024} or post cues in brainstorming meetings to inspire new directions for discussion~\cite{rayan2024exploring}.

While these interjections may offer benefits such as providing useful information to groups or fostering prosocial group dynamics~\cite{doHowShouldAgent2022a, dennisAIAgentsTeam2023}, they can easily become unhelpful or even annoying to group members, who use the same space to communicate with each other~\cite{caiAdvancingKnowledgeTogether2024, avulaEffectsSystemInitiative2022}. 
These systems often abide by rigid message templates and rules defined by system designers, limiting their ability to adapt their content to align with evolving group preferences and specific social contexts~\cite{doErrAIImperfect2023, kimModeratorChatbotDeliberative2021, liu2024compeer}. 
And when they inevitably fail, existing systems often do not provide easy ways for group members to give feedback in the moment~\cite{avulaEffectsSystemInitiative2022}.
As a result, without proper integration into a group's social environment, AI-generated messages can be perceived as irrelevant~\cite{zhang2024ask} or intrusive~\cite{peters2017interrupt}, eventually leading users to ignore the messages or even abandon the system ~\cite{avulaEffectsSystemInitiative2022,rayan2024exploring}.

Recent advances in large language models (LLMs) can generate more flexible and human-like text than templated messages~\cite{yang2024social}. However, these systems often lack the social knowledge required to align their generated content to group preferences. \textit{Retrieval-augmented generation} (RAG) techniques  augment LLMs by incorporating external knowledge that may not be part of the training data~\cite{lewis2020retrieval}. We find that most RAG applications curate and retrieve \textit{factual knowledge} from external sources (e.g., web content, domain-specific documents) to produce fact-based or domain-specific outputs~\cite{kang2023synergi, lee2024paperweaver,kang2024biospark}. Little work to date explores how to augment LLMs with \textit{social knowledge }to generate contextually relevant and socially aligned output in multi-user environments.

In this work, we propose \textbf{Social-RAG}, a workflow for LLM-based AI agents that takes social signals (e.g., topical preferences) gleaned from prior interactions between group members and between members and the AI agent to contextualize and adapt the agent's generation (Figure~\ref{fig:socialRAG}). 
Much like how traditional RAG pipelines use factual knowledge bases to improve factual grounding, Social-RAG retrieves and processes ``social facts'' from a social knowledge base to improve an agent's \textit{social} grounding.
As such knowledge can be difficult to ascertain without tedious user input~\cite{ackerman2000intellectual},
 we instead leverage the rich history of social interactions between group members commonly stored in group spaces and devise strategies for parsing and retrieving meaningful social information from this history.
Group spaces also have many affordances for members to give each other social feedback (e.g., reactions and replies) that can be repurposed for members to give feedback to the agent.

To test Social-RAG in a real-world social setting, we built \textsc{\textbf{PaperPing}}, an AI agent that proactively posts scholarly paper recommendations and LLM-generated explanations into group chat spaces dedicated to research labs or projects.
We chose this test bed due to the collaborative nature of scientific research~\cite{kraut1986relationships, jirotka2013supporting}, where exchanging and discussing scholarly articles is common practice and important for conducting high-quality research~\cite{jirotka2013supporting}; further, rapid  growth in publications motivates researchers' need to identify relevant  papers.  
Indeed, in two formative studies with 39 researchers, we found that researchers frequently recommend papers in group chats 
by targeting a specific member or relating them to prior conversations; they also leverage 
affordances such as emoji reactions and threaded replies to
react to each other's recommendations.
In addition to creating a rich repository for learning people's interests, this existing social behavior also informs \textsc{PaperPing}'s design and how it selects relevant social signals.

\textsc{PaperPing} was deployed for three months in 18 pre-existing Slack channels with a range of social dynamics (e.g., number of participants and frequency of messages), with total exposure to over 500 researchers. 
We find that \textsc{PaperPing} effectively learns researchers' individual and group preferences with minimum upfront effort from users. Users considered its LLM-synthesized messages to be contextually relevant, outperforming generic paper summaries in explaining a paper recommendation to their group. Moreover, \textsc{PaperPing} fits into existing group practices without disrupting group dynamics and fosters the ability of users to reach common ground. 
Our deployment also surfaced areas needing additional investigation, such as how to balance group versus individual preferences. 
We conclude by discussing the broader implications of Social-RAG for designing more adaptive and socially aware AI systems in diverse collaborative settings as well as how to apply our strategies for social grounding in additional contexts.

In summary, our work contributes:
\begin{itemize}[topsep=0pt]
    \item A novel workflow, Social-RAG, for augmenting LLM-based AI agents with social knowledge extracted from past and ongoing group interactions, enabling agents to generate content aligned with the group’s interests and norms.
    \item \textsc{PaperPing}, an AI agent that uses the Social-RAG workflow to proactively post relevant scholarly paper recommendations to a group chat channel with LLM-generated explanations that are contextualized to the group. 
    \item Formative studies with 39 researchers that provide input on how researchers socially interact when discussing academic papers as well as their preferences for an AI agent that recommends papers, which we use to inform the design of \textsc{PaperPing}.
    \item A field deployment of \textsc{PaperPing} on 18 Slack channels reaching over 500 researchers, demonstrating the effectiveness and limitations of Social-RAG in real-world contexts. 
\end{itemize}

\section{Related Work}

\subsection{AI-enabled Systems in Group Social Spaces}

Interactive agents such as chatbots, conversational agents, and virtual assistants have been deployed in group spaces to aid team productivity, providing support on tasks such as conversation summarization~\cite{zhangMakingSenseGroup2018a}, idea generation~\cite{rayan2024exploring}, and collaborative search~\cite{avulaSearchBotsUserEngagement2018}. They also assist in enhancing group interpersonal dynamics, e.g., by supporting conflict resolution~\cite{govers2024ai} or encouraging balanced human participation~\cite{doHowShouldAgent2022a}.
Increasingly, these agents appear as yet another member of a group, interjecting in group conversations to post messages offering facilitation support~\cite{doHowShouldAgent2022a} or sharing information to enhance  decision-making~\cite{zvelebilovaCollectiveAttentionHumanAI2024}.

Despite their benefits, AI interjections in group spaces can be perceived as disruptive due to the timing, relevance, and delivery of their content~\cite{avulaEffectsSystemInitiative2022}, affecting ongoing group discussion~\cite{zvelebilovaCollectiveAttentionHumanAI2024}. Researchers have highlighted the importance of designing AI agents that are socially aware and capable of adapting to the context in which they operate~\cite{rayan2024exploring, heAIFutureCollaborative2024, agarwalConversationalAgentsFacilitate2024, doHowShouldAgent2022a, seeringDyadicInteractionsConsidering2019}.
Yet actually achieving this goal can be challenging due partly to the \textit{sociotechnical gap}, a fundamental issue for computer-supported cooperative work (CSCW) systems that often lack the flexibility to reflect complex social needs~\cite{ackerman2000intellectual}.

Researchers have explored ways to build agents that are sensitive to group dynamics, such as member participation (e.g., speech frequency~\cite{doErrAIImperfect2023}, gaze~\cite{phamEmbodiedMediationGroup2024}) and individual or collective preferences (e.g., opinions expressed~\cite{chiangEnhancingAIAssistedGroup2024, caiAdvancingKnowledgeTogether2024}). Some research shows that agents that demonstrate their awareness of human behavior and learn from human behavior are perceived as more natural and trustworthy~\cite{zhouWordsInfusingConversational2024, leong2024putting, jo2024understanding}, and thus more likely to positively affect team dynamics~\cite{zvelebilovaCollectiveAttentionHumanAI2024}. Similarly, Avula et al.~finds that proactive agents should provide clear justifications for their actions that can demonstrate their social awareness, as  unprompted suggestions may not be immediately understood or valued by group members~\cite{avulaEffectsSystemInitiative2022}. 
However, developing socially aware AI agents remains challenging due to the complexity of human social interactions and the difficulty of fully capturing and responding to them through an AI system~\cite{avulaEffectsSystemInitiative2022, rayan2024exploring}.

\subsection{Strategies for Aligning AI Agents to a Group}

\subsubsection{Direct Elicitation of Group Preferences}
A straightforward method to customize AI agents to a group's context is to allow users to explicitly indicate their preferences to agents. For example, Agarwal et al. developed an agent that polls group members' opinions before sending facilitation messages to the group~\cite{agarwalConversationalAgentsFacilitate2024}. Conversational group recommendation systems also explicitly learn about users' goals through a multi-turn dialog~\cite{vitorioMusicRecommendationSystem2023a, nguyenConversationalGroupRecommender2017, nguyenDynamicElicitationUser2017, nguyenChatBasedGroupRecommender2017}. Studies that use this method show that continuous and interactive feedback mechanisms lead to better alignment with user expectations~\cite{guzziInteractiveMultipartyCritiquing2011,contrerasIntegratingCollaborationLeadership2021a}.

However, explicit feedback often requires users to invest prohibitively high upfront effort, exclusively for the purpose of informing systems about their preferences. As a result, users see the benefits of the system only after substantial effort, making systems suffer from a 'cold start problem' and hindering their adoption~\cite{nguyenDynamicElicitationUser2017}. Moreover, frequent interactions in group spaces that are solely for steering the system could potentially disrupt group dynamics (e.g., a too frequent poll prompt)~\cite{agarwalConversationalAgentsFacilitate2024}.

\subsubsection{Learning Group Preferences from Activity Traces}
A unique asset of real-world groups is their rich group communication history, which captures the process of individual and team activities. As a result, researchers have explored using this history to make inferences about individual and group preferences~\cite{shamiMakingSenseStrangers2009} and to inform interactive systems that facilitate team processes, such as peer evaluation~\cite{shiValueActivityTraces2023b} and information summarization~\cite{zhangMakingSenseGroup2018}. Researchers have also explored the design of AI systems that leverage human social interaction. For instance, Krishna et al. showed that collective activity traces of whether users respond to agents' questions can teach an AI agent to ask socially appropriate questions~\cite{krishnaSociallySituatedArtificial2022}. Others have explored providing agents with contextual information about a group gathered from documents and ongoing conversations to facilitate social awareness~\cite{parkCoExplorerTechnologyProbe2024, caiAdvancingKnowledgeTogether2024, chiangEnhancingAIAssistedGroup2024}. 

Despite their potential, many current strategies for contextualizing agents were validated only in controlled lab environments~\cite{doHowShouldAgent2022a, avulaEmbeddingSearchConversational2019, avulaEffectsSystemInitiative2022} or using a Wizard-of-Oz approach~\cite{ringAddressingLonelinessIsolation2013a, rayan2024exploring}. Researchers have yet to explore how AI agents can continuously monitor group interactions to adapt their messages to suit the preferences of diverse teams in the wild. Our work takes this next step.

\subsubsection{Designing AI Agent Behavior to Fit Social Norms}
Many existing agents are developed with rule-based algorithms or pre-defined language templates that aim to universally apply to different deployment contexts as opposed to aligning to the social norms of that context~\cite{doHowShouldAgent2022a, avulaEmbeddingSearchConversational2019, avulaEffectsSystemInitiative2022}. 
As a result, agents can struggle to convince users of their relevance and utility~\cite{liu2024compeer}.
The capabilities of large language models (LLMs) for flexibly generating human-understandable text~\cite{park2023generative} open up new potential for agents to integrate more naturally into multi-user shared spaces. Researchers have explored using LLMs to enable personalized interactions that cater to different individuals. For example, Ha et al.~\cite{haCloChatUnderstandingHow2024} created personalized agent personas with distinct expertise to increase the relevance of the information shared with  users. Lian et al.~\cite{lianRecAILeveragingLarge2024} explored the potential for LLMs to act as personalized recommenders and to tailor explanations to individual user needs.

Despite this promise, integrating LLM agents into practical multi-user systems remains a challenge. Traditional LLM agents are trained or fine-tuned on datasets representing general world knowledge, which are not necessarily tailored to the preferences and backgrounds of specific groups. In fact, current LLMs predominantly focus on informational content instead of social factors~\cite{Hovy2021TheIO} and thus show limited abilities to reason about social knowledge (e.g., speaker intention, social proximity of speaker and receiver)~\cite{Choi2023DoLU, Li2024SocialID}.
Moreover, natural language processing (NLP) research has shown that LLMs often lack social intelligence (i.e., cognitive intelligence, situational intelligence, and behavioral intelligence)~\cite{yang2024social}. 
These factors limit their capability to understand and interpret the dynamics of a group's conversation and subsequently generate responses that are natural to the group's context.

\subsubsection{Retrieval Augmented Generation (RAG)} 
One way to augment LLM generation with knowledge not in the training data is via retrieval augmented generation (RAG)~\cite{lewis2020retrieval}. RAG was originally proposed to address the issue of LLMs lacking access to a constantly expanding and changing body of knowledge, leading to false information (``hallucinations'') in the context of NLP tasks such as question answering~\cite{shuster2021retrieval,peng2023check,mallen2023not,vu2023freshllms, lewis2020retrieval} and fact checking~\cite{zhou2024muse,yue2024evidence}.
Therefore, most RAG systems are typically built using web data~\cite{zhou2024muse,vu2023freshllms,gao2023rarr} or pre-compiled static document corpora or knowledge bases, such as Wikipedia~\cite{shuster2021retrieval} or a domain-specific corpora~\cite{yue2024evidence, kazemitabaar2024codeaid, yen2024search, kang2023synergi, kang2024biospark}. Leveraging information retrieval (IR) techniques such as BM25~\cite{robertson2009probabilistic} or semantic embeddings~\cite{yang2024aqua,lewis2020retrieval, lee2023dapie, pennington2014glove}, these systems retrieve relevant documents to include in their prompts. A plethora of recent human-computer interaction (HCI) research leverages RAG to build interactive systems that more effectively support knowledge-intensive tasks, such as sensemaking~\cite{liu2024selenite}, decision-making~\cite{kazemitabaar2024codeaid, yen2024search}, or brainstorming~\cite{kang2023synergi, kang2024biospark}. However, scant research has explored how RAG can be used towards addressing the sociotechnical gap.

More closely related to our work, research has used RAG to personalize text generation by retrieving data from personal interaction histories~\cite{hou2024my, liu2024unblind, rajan2024arizona, zerhoudi2024personarag, sayana2024beyond}. However, these applications have been limited to individual users and often require explicit, text-based cues from past interactions to infer user needs~\cite{rajan2024arizona, liu2024unblind, park2023generative}. In contrast, in group contexts, social knowledge is often implicitly embedded in observable social interactions (e.g., emoji reactions to messages) because explicit preference elicitation can be disruptive in shared spaces~\cite{agarwalConversationalAgentsFacilitate2024}.
As a result, additional inference is needed to interpret such implicit signals, and existing RAG techniques, such as embedding textual messages, retrieving based on textual similarity, or directly injecting group messages into prompt contexts, cannot capture and apply social knowledge effectively. 

In contrast to prior work and informed by our study of specific user needs in a group setting, this paper proposes to (1) dynamically curate a database of \emph{social facts} learned from past and ongoing interactions in an online group and (2) leverage RAG techniques to build an AI agent that can generate socially grounded messages to interact with group members.

\section{Social-RAG: Socially Grounding AI Generation via Retrieval  from Group Interactions}

In the following, we describe \textit{design goals} for our workflow that are drawn from prior work and through building, iterating on, and using a prototype AI agent (described in more detail in Section~\ref{sec:system}).
We then describe the \textit{steps of our workflow}, including the curation of a social knowledge base gleaned from group interactions and the retrieval pipeline for real-time zero-shot learning for message generation.

\subsection{Design Goals}

Our design goals are informed by prior literature as well as a \textit{research through design} (RtD) approach~\cite{zimmerman2007research}, where we iteratively designed and built a prototype LLM-based AI agent over the course of 18 months that we deployed to solicit user feedback in our own social spaces. In Section~\ref{sec:system}, we provide details of our final system and evaluation. Though we built the system for a specific task and social context, the process of iteratively working through design requirements helped us gain insights into the broader question of how to best leverage existing group social information and infrastructure to align an AI agent to a group.

 \textbf{DG1: Contextualize suggestions in the group's social history.}
Research has shown that tailored explanations that convey the relevance of AI suggestions~\cite{liaoWhatCanYou2016, ehsan2021expanding} and citations to past social interactions~\cite{cox2023comparing} can improve users' trust in and acceptance of AI agents. With access to a group's social history, we can use the same approach to contextualize AI suggestions to the group.

\textbf{DG2: Integrate AI content into the group without disturbing existing group social practices.}
Although the AI agent will be interacting with a group in its shared social space, it should not crowd out existing productive human-human interactions (e.g., by posting overly long messages).
Poor integration could result in an undesirable group outcome, such as discouraging discussion or making users leave for other channels.

 \textbf{DG3: Minimize upfront user effort for specifying  preferences to the AI and also enable continued improvement.}
Many existing AI agents require explicit 
preference choices from users (e.g., by polling users' opinions~\cite{agarwalConversationalAgentsFacilitate2024}), which can be tedious.
We can learn from prior group interaction data to determine preferences, reducing user effort needed to bootstrap the agent. We can also lower barriers to user feedback for additional adaptation by leveraging the existing affordances of the space and existing common behaviors of users.

 \textbf{DG4: Promote transparency of information shared between humans and the agent to maintain common ground in the group.
} 
Maintaining common ground is essential for productive asynchronous collaboration~\cite{olson2000distance}.
Thus, agent communication to the group, even when targeted to specific members, should occur in the shared group space. Users' feedback to the agent should also be visible to others to ensure transparency and help group members learn and respect each other's preferences.

\subsection{Social-RAG Workflow}

We now describe the steps of the Social-RAG workflow (Figure~\ref{fig:socialRAG}) in detail and explain how they instantiate our four design goals.

\subsubsection{Step 1: Collect and index content from group interactions to create a social knowledge base}
The first step involves determining how to leverage a group's pre-existing interaction history to populate a social knowledge base, minimizing upfront user effort (DG3).
Depending on the specific goal of the AI agent's interjection and technical feasibility, the scope of relevant items in the history could vary. 
For instance, given a bot that recommends the latest news items to a group that likes to discuss the news, prior messages discussing news-related topics and news articles posted by group members could be collected as an indication of group members' interests. 
In addition, group members' reactions and engagement around these messages could signal the preferences of more group members. 
For example, if a group chat had an active discussion about an article, then it is possible that another article on a similar topic could inspire future conversations.

After collecting relevant posts, they must be parsed to extract indexable items for an algorithm to retrieve since we found that simply providing snippets of raw conversation history as context to an LLM performed poorly. Items to be extracted could include any relevant information for suggestions, such as an article URL for a bot that recommends articles. 
How to index an item depends on what aspects of the item should be used to determine relevance. For instance, news articles could be indexed by the semantic embedding of the article title.

Next, each item is associated with additional information.
Each item should initially be associated with the \textbf{posts} that mention it.
Additional context, such as the sentiment of the message or reactions to it, could also be stored to better identify the \textbf{direction of preference}, as opposed to simply interest.
Additional \textbf{metadata} about each item could be gathered from external data sources or through additional analysis (e.g., the publisher of an article or its main topics).
Indexed items can also be associated with the  \textbf{group members} that mention or interact with the entity so that the agent knows who in the group would be most interested. 

This index can now be used to improve the underlying suggestions that the AI agent provides to better target the group.
For instance, when recommending a new article to the group, a news recommendation bot could query the index to determine whether the article might be interesting to the group (i.e., if the group has discussed similar articles) and recommend only those articles that return the most results.

\subsubsection{Step 2: Retrieve social signals from knowledge base and contextualize AI explanations}
\label{sec:socialsignals}

After the agent decides on a suggestion to post,
it can use the social knowledge base to help explain the relevance of the suggestion to the group (DG1). 
We present three categories of social signals that can be retrieved and how those signals contextualize AI suggestions in different ways:

\textbf{Cite relevant prior posts in group history.}
Given an index of items previously mentioned and the posts from which they came, the agent can retrieve a \textbf{prior post }discussing similar items. The agent could cite this prior message and its associated thread of discussion when suggesting a new item to explain its relevance. The citation could then link to the prior post, enabling members to investigate the connection and establish trust in the agent's suggestions; 
this is akin to how traditional RAG systems cite the sources from which they retrieved information in their generation.
The ability to see older relevant discussions also reminds group members of what was already discussed to further establish common ground (DG4) for continuing discussion.
For instance, an AI agent that makes automated \textit{pull requests} (PRs) in a code repository with suggested code edits could link to prior PRs that recently edited the same piece of code or is attempting to fix the same problem.

\textbf{Highlight relevant metadata in an explanation.} Each item in our index can be associated with one or more pieces of \textbf{metadata} gathered from external data sources. 
From this information, the agent can determine what metadata is potentially interesting to the group by storing and retrieving counts of how often an item with that metadata was mentioned.
For instance, if group members in the past have shared many news articles from a particular writer or publisher, they might be more interested in a new article if they know that it is also from that writer or publisher.
Thus, when a new item is set to be suggested to the group, the agent can also determine whether any particular metadata about it should be mentioned when explaining the suggestion.

 \textbf{Mention relevant group members.}
Finally, each item in our index can be associated with one or more specific \textbf{group members} who posted about the item or reacted or responded to a post about that item. 
When a new item is set to be suggested, the agent can determine as a signal of interest which items it is most similar to and which group members interacted with those items.
Additionally, it could use external data sources with other information about group members to determine connections to specific group members.
Then, the AI agent can @-mention specific group members as another explanation of relevance when posting to the group. 
Alerting all group members of specific members' interests could potentially encourage them to interact with the agent's message in the channel and make them aware of each other's interests (DG4). 

\subsubsection{Step 3: Rank social signals and generate a succinct message for the AI agent to post}
In this step, 
we gather all relevant social signals associated with a suggestion and combine them into a message for the agent to post. Our goal is to craft messages that naturally suit the existing human-human conversational  cadence in the social space (DG2). To achieve this goal, we leverage LLMs (1) to flexibly generate a message in natural language mimicking existing group dialogue and (2) to summarize social signals in a succinct way to avoid overwhelming group members.
We take the social signals gathered in the last step and the original suggested item and provide them as context to an LLM, along with instructions and examples of how the AI agent should provide its suggestion and explanation. 
For instance, the prompt could ask for a specific format, length, or tone. 
If the prior step resulted in many social signals retrieved, 
we can use heuristics to select one or a few most relevant signals to provide as context or simply provide all information to the LLM and let it determine what to retain to form a short explanation.

\subsubsection{Step 4: Post generated message to social channel and collect and index group reactions}
Finally, with an LLM generation in hand, the AI agent can post the message in the group channel for all members to see (DG4). At this point, as with any other human-generated post, members can interact with the post, such as by reacting or replying to it.
These interactions can also be collected and indexed (similar to how group member posts are collected and indexed in Step 1).
Not only does enabling group reactions to the post serve as a way to further learn about other group members' preferences (DG4), but by leveraging the existing affordances of the social space, we provide an easy and more natural way to give feedback to the AI agent for additional adaptation (DG3).

\section{\textsc{The PaperPing} System}~\label{sec:system}

We implement the Social-RAG workflow in a system called \textsc{PaperPing}, which automates academic paper recommendations in a scholarly group chat context.
We chose this as our test setting due to the highly collaborative nature of scientific research, which often involves regular, information-dense communication among collaborators~\cite{krautPatternsContactCommunication1988, paine2017has, kraut1986relationships, jirotka2013supporting}, much of which occurs today using digital tools. 
We also chose to focus on paper recommendations since academic papers play an important role in facilitating knowledge sharing, fostering collaboration, and informing decision making in scientific team  communication. 
The process of discovering and searching through papers is often open ended and opportunistic~\cite{nedumovExploratorySearchScientific2019}. Finally, paper recommendations often come from social networks~\cite{soliman2024mitigating}, and researchers commonly form groups for the purpose of sharing and exchanging relevant research literature.

\subsection{Formative Studies}

We first describe two formative studies we conducted with scholars to identify (1) if and how they use social channels to discuss and recommend papers to each other, and (2) their preferences for an automated paper recommendation bot. Findings from these studies motivate and inform the system design of \textsc{PaperPing}, including the specific social signals (e.g., paper metadata) we choose to capture about each paper and the design of automated messages (e.g., format, tone).
 Our study was reviewed and deemed exempt by our university's Institutional Review Board (IRB).

\subsubsection{Procedure and Participants}
In our first study, we \textbf{interviewed }13 researchers who regularly discuss research in group chat channels. The participants were part of research project and interest groups, such as a lab discussion channel, focused mainly in the areas of HCI and NLP. The interviews lasted 19 minutes to 1 hour, and we asked participants about their motivations and considerations when sharing scientific literature and their experiences with  literature recommendations. We analyzed the interview transcripts using inductive coding and clustered the codes into common themes. 

In the second study, we conducted a \textbf{survey} with 26 participants from two active research discussion channels on Slack, consisting of a mix of research scientists, professors, PhD students, and undergraduate students. In the survey, we asked participants to compare different styles of Slack messages that recommend academic papers in order to surface people's preferred message presentation in an academic Slack setting.
These styles varied across several dimensions, including tone (neutral vs. enthusiastic, promotional language, humor), use of emojis, and formatting (single block of text vs. bulleted format, use of text highlighting such as bold). Finally, we asked participants to indicate their preferences for pairs of messages showcasing different styles. 

\subsubsection{Results} 

\textbf{Researchers regularly share papers with each other in group chat}.
In project-specific channels, interviewees share papers that directly support specific goals, such as answering research method questions or illustrating ideas to team members. 
These shared papers play a crucial role in shaping the research project by providing potential citations, serving as model papers, and enhancing collaboration by ensuring consensus of perspective.  
In interest-based channels, papers are shared for their general relevance to the group's common interests; although not shared in response to specific questions, these papers spark new conversations and facilitate relationship building among like-minded researchers.

\begin{figure}[h]
    \centering
    \includegraphics[width=0.45\textwidth]{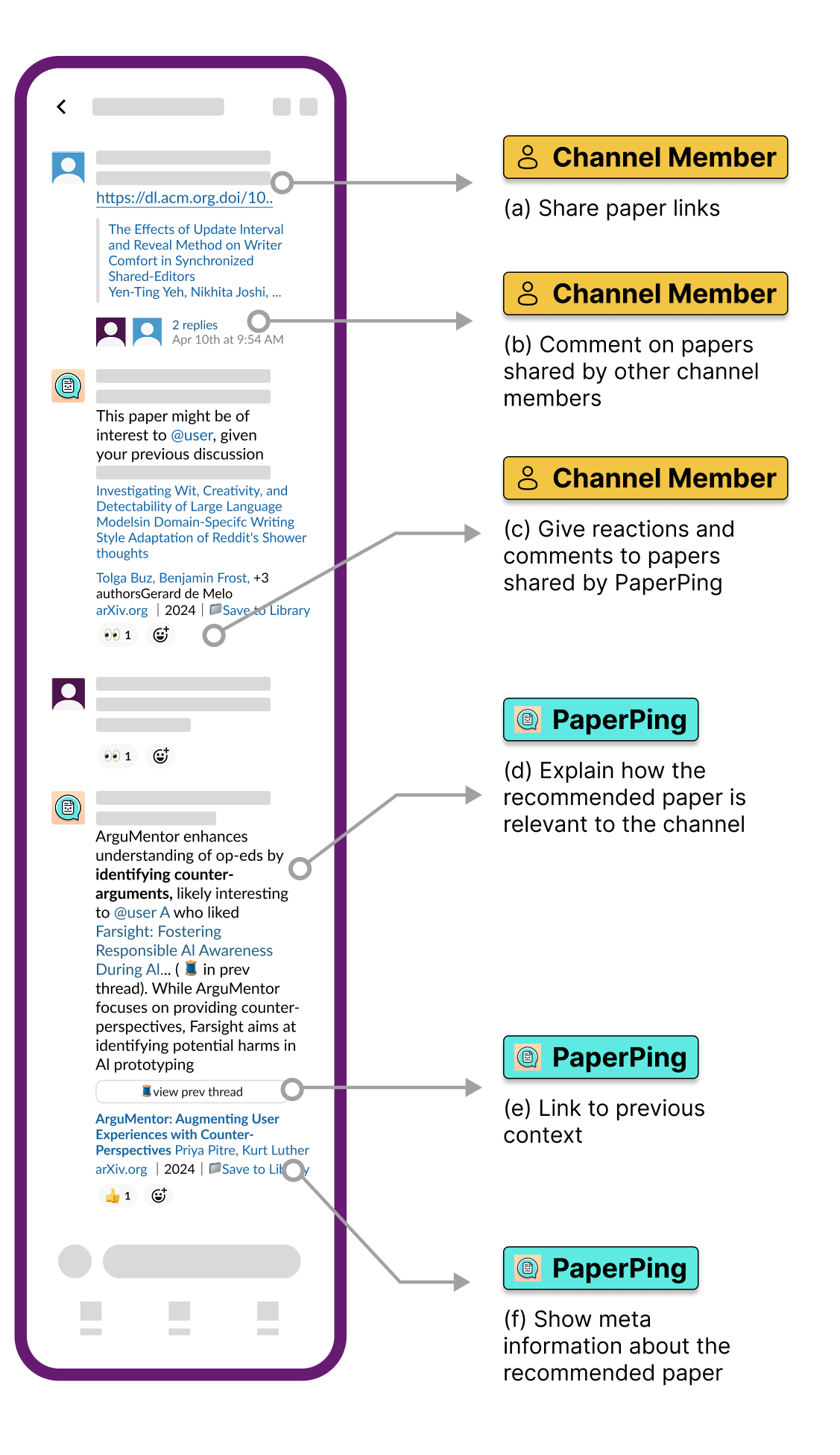}
    \caption{\textbf{A demonstration of  how users interact with \textsc{PaperPing}. }Channel members share paper links, react to papers shared by other members or \textsc{PaperPing} and comment on shared papers. After gathering and processing this information, \textsc{PaperPing} sends new contextually grounded paper recommendations, with explanations of how the recommended papers are relevant to the channel. It also provides links to previous related discussions and includes meta-information about the recommended papers in the recommendation message.}
    \Description{
    This figure demonstrates how \textsc{PaperPing} interacts within an example user scenario. Channel members share paper links, react to papers shared by other members or \textsc{PaperPing}, and comment on shared papers. After gathering and processing this information, \textsc{PaperPing} sends new contextually grounded paper recommendations, including explanations of how the recommended papers are relevant to the channel. It also provides links to previous related discussions and includes meta-information about the recommended papers in the recommendation message.}
    \label{fig:PaperPingScenario}
\end{figure}

\textbf{When sharing papers, researchers attempt to connect the papers with group context}.
Interviewees who post paper recommendations mentioned that they like to contextualize the recommendation in the discussion. For example, P1 often highlights specific quotes or explains reasons for sharing the paper. 
They also use @-mentions to draw the attention of relevant individuals to specific papers. 
Interviewees looked at prior papers that were shared and group engagement about them to determine what kinds of papers to share, how to share, and the frequency of sharing.
However, some said it currently takes too much effort to craft a message explaining a paper recommendation, calling it ``\textit{free labor.}'' 
They are occasionally hesitant to write their own paper summaries since they often recommend papers without having fully read them and worry about providing an inaccurate summary.   
Some junior scholars also worry about interrupting group dynamics or taking too much space with papers that may be perceived to be low quality or irrelevant to their groups.

\begin{figure*}[t]
    \centering
    \includegraphics[width=\textwidth]{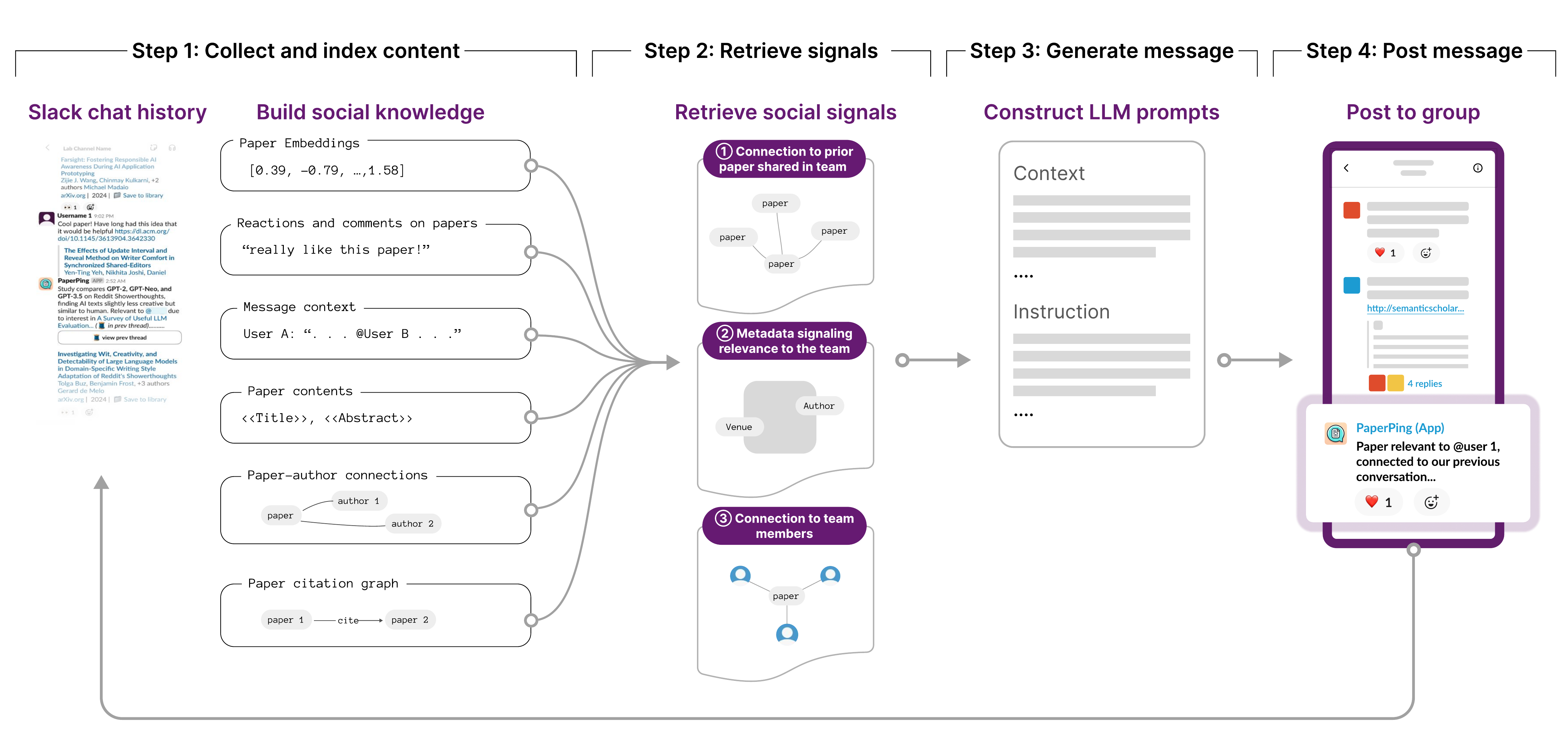}
    \caption{\textbf{The \textsc{PaperPing} implementation, leveraging the Social RAG workflow.} \textsc{PaperPing} extracts relevant social information from two data sources: the Semantic Scholar knowledge base, providing paper content, paper-author connections, and citation graphs; and Slack chat history, which includes previously mentioned papers, emoji reactions, comments, and user interactions (Step 1). \textsc{PaperPing} then retrieves relevant social signals from the data using three heuristics (relevant metadata, the most similar and discussed prior paper in the chat, and the member most likely to be interested) developed over a three-month iterative design process (Step 2). Next, \textsc{PaperPing} uses chains of LLM prompts to turn the retrieved social signals into a natural language explanation (Step 3). Finally, \textsc{PaperPing} posts the message to research group channels, where group members can provide emoji reactions or reply to the message to give feedback (Step 4).
    }
    \Description{Figure 3: This is an architecture diagram of  PaperPing's implementation, guided by the Social-RAG pipeline. The Social RAG pipeline in PaperPing includes two data sources: the Semantic Scholar knowledge base, providing paper contents, paper-author connections, and citation graphs; and Slack chat history, which includes previously mentioned papers, emoji reactions, comments, and user interactions. These data sources are used to build social knowledge, whose indexes include Paper Metadata API, Mentioned Paper Index, and Member Preference Index. This is Step 1, collect and index content in the Social-RAG pipeline.
The Step 2: retrieve signals involve retrieving social signals based on three heuristics: connection to prior paper shared in team, metadata signalling relevance to the team, and connection to team member.
For prompt construction process (Step 3 generate message), given a paper, PaperPing retrieves relevant metadata, the most similar and discussed prior paper in the chat, and identifies the member most likely to be interested. Then, these signals are ranked according to importance heuristics, and finally, an explanation is generated and posted to the slack group (Step 4 post message).}
    \label{fig:PaperPingSys}
\end{figure*}

\textbf{Researchers reading papers recommendations prefer short and neutral explanations}.
Survey participants overwhelmingly preferred straightforward and neutral Slack messages when introducing academic papers in their channels. Messages with minimal formatting and without text decoration or humor were favored, as seen in the strong preference for neutral tone (92.3\%) and simple formatting (96.2\%). 
Promotional language and humor were particularly unpopular, with only 15.4\% and 7.7\% of participants supporting those styles, respectively. While there was some division on emoji usage---undergraduates were more receptive (60\%) compared to PhD-level participants (43\%)---the overall trend leaned toward simplicity and clarity.

\textbf{Researchers prefer quick ways to acknowledge paper recommendations.}
When responding to papers that others shared in chat, interviewees commonly use emojis and quick comments to indicate their interest as a way to show appreciation and encourage others to continue to share. Emojis are liked because they are ``\textit{unobtrusive and don’t take space}.''

\subsubsection{\textbf{Summary}}
Our findings motivate the utility of a paper recommendation bot that posts in group chat since researchers already perform this activity but occasionally find it difficult to do so easily and succinctly. In terms of bot  design, interviewees provide examples of ways they connect their recommendation to the group; these inform the social signals we implement into \textsc{PaperPing} from the Social-RAG workflow. In addition, our findings guide our effort to design bot interjections that are short and unobtrusive and to give feedback to the bot that is quick and easy.

\subsection{\textsc{PaperPing} Example User Scenario}

Figure~\ref{fig:PaperPingScenario} demonstrates how users interact with \textsc{PaperPing}. As an example scenario, imagine that
Kara leads an interdisciplinary research group consisting of PhD students, postdocs, and undergraduate researchers. The group also shares common interests. To help the group keep current with the latest research, Kara encourages everyone to share and discuss relevant papers in their group chat on Slack. Upon learning about \textsc{PaperPing}, Kara introduces it to her group to help them stay informed about the latest literature without adding to their workload. She set the recommendation frequency to be once every other day.

One day, \textsc{PaperPing} notifies the group of a new paper recommendation. The message includes a brief explanation that highlights the connection between the recommended paper and their recent chat, citing a previous discussion thread for reference. The message is also personalized with an @-mention to her postdoc Ryan since he has previously reacted positively to similar papers. 
Ryan views the linked previous discussion to gain more context and has some new thoughts about a question that confused the group in previous discussions. He shares his thoughts under the latest paper recommendation while @-mentioning other relevant students and up-voting the bot's post.
Kara sees the discussion and contributes another paper that she recently encountered and thought relevant. 
The team continues to interact with \textsc{PaperPing} and with each other as recommendations arrive. 
Over time, the positive feedback and ongoing interactions from the group help \textsc{PaperPing} refine its knowledge of the team’s preferences, recommending more relevant papers and providing more targeted explanations.

\subsection{\textsc{PaperPing} Process and Features}

Building on findings from our formative interviews and survey, we describe in more detail \textsc{PaperPing}'s process and features specific to our academic paper recommendation use case (Figure ~\ref{fig:PaperPingSys}).

\begin{figure*}[t]
    \centering
    \hspace*{-1cm}
    \includegraphics[width=1.1\textwidth]{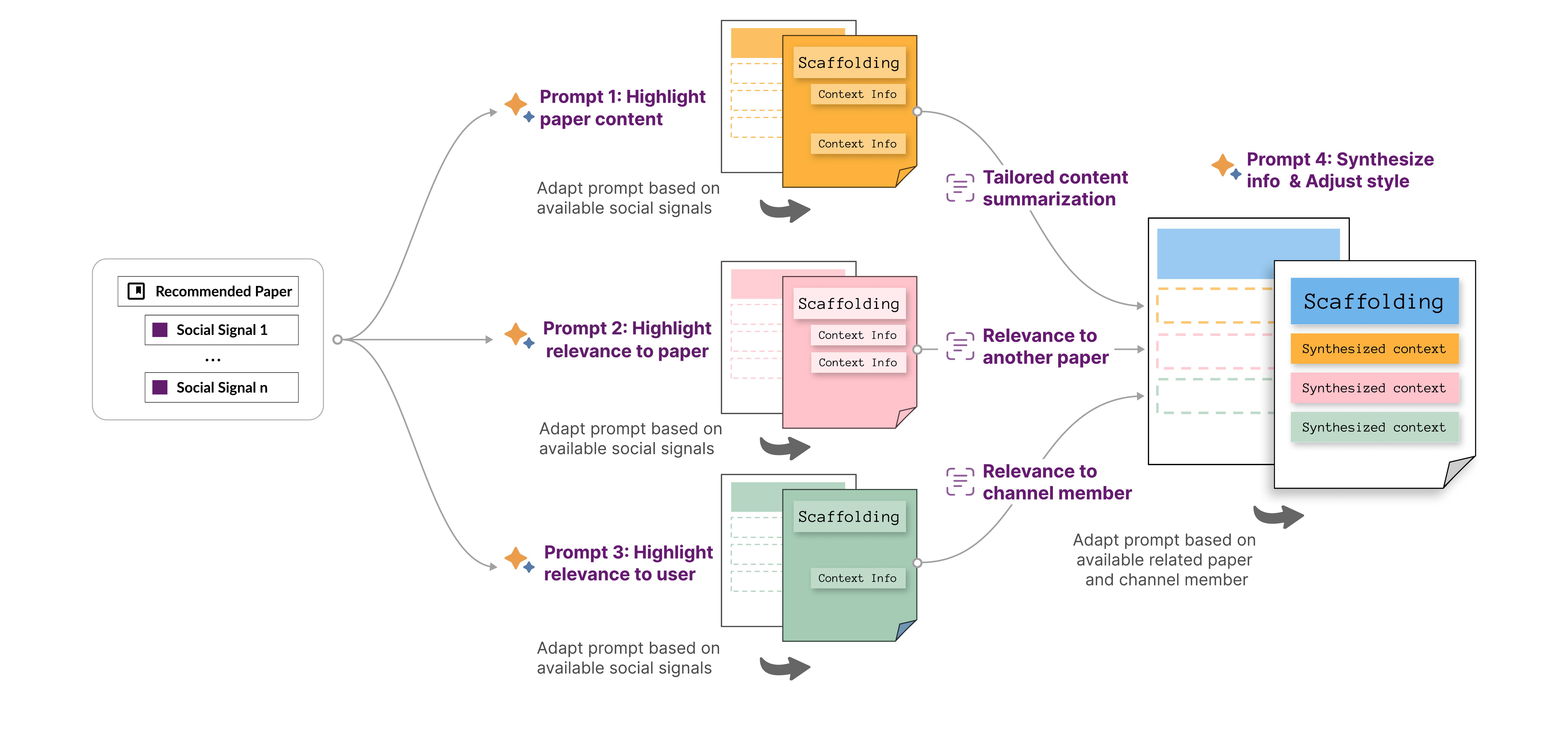}
    \caption{\textbf{\textsc{The PaperPing} prompts pipeline (Condition 4)}. The pipeline starts with social signals leading to three prompts: Prompt 1 highlights paper content; Prompt 2 highlights relevance to a previous paper; and Prompt 3 highlights relevance to a channel member. Finally, outputs from these prompts feed into Prompt 4, which synthesizes the information and adjusts the style, resulting in the final output.}
    \label{fig:prompts}
    \Description{Diagram showing \textsc{PaperPing} prompts pipeline for prompt 4. It starts with social signals leading to three prompts: Prompt 1 highlights paper content, Prompt 2 highlights relevance to a previous paper, and Prompt 3 highlights relevance to a channel member. Finally, outputs from these prompts feed into Prompt 4, which synthesizes the information and adjusts the style, resulting in a final output.}
    
\end{figure*}

\subsubsection{Step 1}

To create a social knowledge base from prior chat history, \textsc{PaperPing} parses prior messages to extract links to academic papers shared by others. It converts these links into an embedding for each paper to store in a vector database, along with the messages mentioning that paper, message authors, and when messages were posted. In addition, it collects users' emoji reactions to the message, noting which emojis have a positive versus negative sentiment (e.g., thumbs up as a positive reaction) and any comments in the thread under the message.

 \textsc{PaperPing}  also links this paper database with external data sources that provide additional relevant metadata. In particular, it collects author names on the paper, their affiliations, the papers they cite and that cite them, and the venue in which the paper is published.
Finally, \textsc{PaperPing} optionally allows users to link their group chat account to their academic publication record, letting it know the user's affiliations and publication history.

When \textsc{PaperPing} prepares to make a recommendation, it sends to an off-the-shelf paper recommender system recent items from this paper database that have (1) received positive emoji reactions, or (2) one or more comments in response from a group member.

\subsubsection{Step 2}

After accessing a paper, \textsc{PaperPing} next retrieves social signals related to it.
It focuses on three groups of social signals: 
(1) Connections from the new paper to previously shared papers in the chat channel, looking for potential topic overlaps, citation relationships, or common authorships. 
(2) Metadata items about the paper, such as authors or publisher, that have some connection to prior interactions in the chat channel; 
for example, if an author of the new paper was previously mentioned in the group chat history, it will identify the author's name as a social signal. 
(3) Team members who might be interested in the paper based on their previous interactions in the chat channel. 
These align with the three types of social signals discussed in Section~\ref{sec:socialsignals}.
If it retrieves many social signals from this step, PaperPing uses heuristics developed over several months of piloting to rank and select only a few that it deems the most important. See the Appendix for a detailed list of retrievable social signals implemented in \textsc{PaperPing}.

\subsubsection{Step 3}
As Figure~\ref{fig:prompts} shows, \textsc{PaperPing} uses chains of LLM prompts to turn the retrieved information into a natural language explanation. The prompt chains are dynamically constructed based on the availability of social signals from Step 2. 
\textsc{PaperPing} uses one prompt for each group of social signals before using one final prompt to succinctly summarize the information and implement a friendly and informative tone. 
This approach lets the output address meaningful connections between the recommended paper and the channel context while reducing hallucinations. We iteratively developed these prompts by piloting our prototype in two channels we participate in that are not part of the deployment study. The Appendix provides the specific LLM prompts we used.

\begin{table*}[h]
\small
\begin{tabular}{|r|l|r|l|r|l|}
\toprule
\multicolumn{1}{|l|}{\textbf{Index}} & \textbf{Channel Type}    & \textbf{Channel Size} & \textbf{Frequency of Recommendation}         & \multicolumn{1}{l|}{\textbf{Study Duration (weeks)}} & \textbf{Participants}  \\ \midrule
1                           & Lab channel     & 31-100       & every-other-day                     & 14& P1, P2, P4    \\ \hline
2                           & Common interest & 2-10         & every-other-day                     & 14& P3, P5        \\ \hline
3                           & Lab channel     & 100+         & weekly                              & 21& P12, P17      \\ \hline
4                           & Lab channel     & 31-100       & weekly                              & 18                                    & P10, P15      \\ \hline
5                           & Project         & 2-10         & every-other-day                     & 24                                    & P8, P23       \\ \hline
6                           & Project         & 2-10         & every-other-day                     & 24                                    & P8, P23       \\ \hline
7                           & Lab channel     & 10-30        & every-other-day                     & 24                                    & P8, P23       \\ \hline
8                           & Project         & 2-10         & every-other-day                     & 24                                    & P8, P23       \\ \hline
9                           & Lab channel     & 2-10         & every-other-day                     & 24                                    & P14, P20      \\ \hline
10                          & Project         & 2-10         & daily                               & 23                                    & P11           \\ \hline
11                          & Seminar         & 31-100       & weekly                              & 16                                    & P16, P9       \\ \hline
12                          & Seminar         & 31-100       & every-other-day                     & 23                                    & P13, P18, P19 \\ \hline
13                          & Common interest & 10-30        & weekly                              & 10                                    & P24           \\ \hline
14                          & Lab channel     & 10-30        & weekly                              & 15                                    & P24           \\ \hline
15                          & Lab channel     & 10-30        & daily                               & 2                                     & P6, P7         \\ \hline
16                          & Common interest & 10-30        & weekly                              & 23& P21. P25*     \\ \hline
17                          & Common interest & 10-30        & weekly                              & 24                                    & P26*, P27*    \\ \hline
18                          & Common interest & 10-30        & every-other-day-\textgreater weekly & 22                                    & P22, P28*     \\ \bottomrule

\end{tabular}

\vspace{0.3cm} 

\caption{\textbf{Participant information by channel type and recommendation frequency.} Participants marked with an asterisk (*) participated asynchronously in the offline evaluation task without joining the exit interview. Duration counts the time from the installation of \textsc{PaperPing} until all participants in the channel had completed the study. While not counted in the study duration, 11 channels continued to use \textsc{PaperPing} voluntarily for more than six months after the study concluded.}
    \label{tab: participants_info}
\end{table*}

Our LLM prompt set instructs the generation to take the following format. The generation first introduces the paper using a contextualized explanation based on the preceding social signals. Whenever the explanation mentions a user, it will also @-mention the user, unless it has already done so recently. 
When there is a relevant previously mentioned paper, 
the explanation also cites the previous thread that mentions the paper; this citation is linked, allowing group members to easily view the previous thread.
Finally, the explanation also provides the paper's meta information, such as title, author name, venue, etc., in a structured format separate from the contextualized explanation.
The text design came from feedback from formative study participants and pilot users. For example, users indicated that they prefer to see keywords bolded in the explanation to help them skim through the text. 

\subsubsection{Step 4}
The message gets posted in the group channel, after which group members can provide an emoji reaction or reply to the message to give feedback.
Group members can adjust the frequency of recommendations to daily, once every other day, or weekly based on their individual preference.

\subsection{\textsc{PaperPing} System Implementation}
\textsc{PaperPing} is implemented as a Slack application, with recommendations arriving in the group conversation as chat messages from the \textsc{PaperPing} bot account. It is built with the Slack Web API. The backend server is built in Node.js and interfaces with a PostgresSQL database. \textsc{PaperPing} leverages the Semantic Scholar Recommendation API\footnote{\url{https://api.semanticscholar.org/api-docs/recommendations}} to select paper recommendations based on a group's interests and prior interactions. To retrieve the paper's metadata, SPECTER embeddings for each paper~\cite{cohan2020specter}, and citations from the citation graph, \textsc{PaperPing} uses the public Semantic Scholar Academic Graph API. Finally, \textsc{PaperPing} uses \texttt{GPT4-turbo-preview} through the OpenAI API for all text generation with an LLM.

\section{Field Deployment}
We conducted a \textbf{three-month field study} where we deployed \textsc{\textsc{PaperPing}} in 18 Slack channels, collectively reaching over 500 channel members. From these channels, we collected in-depth feedback from 28 members via \textbf{exit interviews and questionnaires} and an \textbf{offline evaluation task}. Three participants (P8, P23, P24) were members of multiple channels that installed \textsc{PaperPing}, while all others were in only one participating channel. All but 
one channel installed \textsc{PaperPing} for at least 3 months; members of that channel, who decided 
to discontinue the study after one week, were still invited to debrief about their experience, and their feedback is also reported below. Eleven channels continued to use \textsc{PaperPing} voluntarily for more than six months after the study concluded. The study was reviewed and deemed exempt by our university's IRB.

\begin{figure*}[h]
    \centering
    \includegraphics[width=\textwidth]{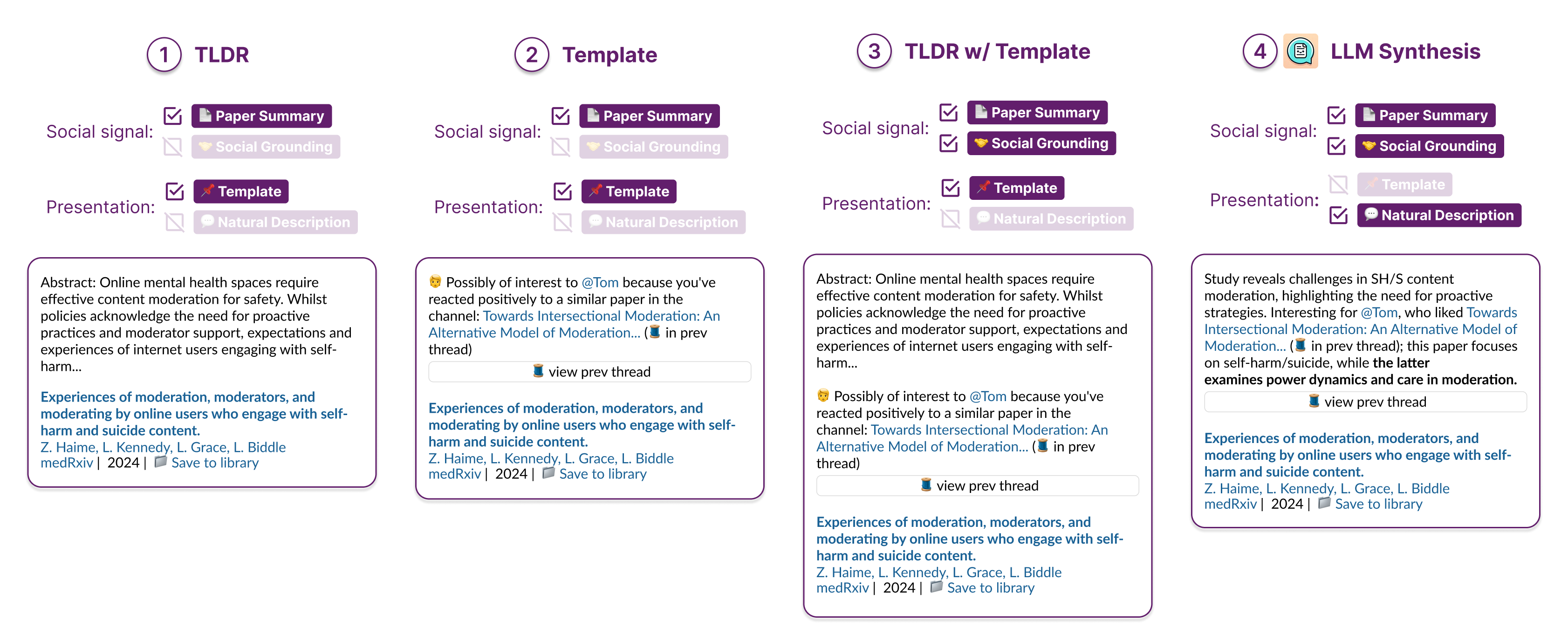}
    \Description{
    This figure shows the four different example messages for the \textsc{PaperPing}. It first provides a table describes the four different variations of \textsc{PaperPing}.  Then, it follows the four real example messages from the field study.
    }
    \caption{\textbf{Four \textsc{PaperPing} conditions shown in the offline evaluation task and example explanations.} Names are anonymized.}
    \label{fig:condition_figures}
\end{figure*}

\subsection{Research Questions}
Our research questions are based on our designs goals for Social-RAG (Section 3.1).
We focus on identifying how channel members perceive and interact with \textsc{PaperPing}'s suggestions and how \textsc{PaperPing} affects group dynamics. In particular, our research questions include:

\begin{itemize}
    \item RQ1: How effective and effortful are \textsc{PaperPing}'s feedback mechanisms in eliciting individual and group interests? How relevant are \textsc{PaperPing}'s suggestions? (DG3)
    \item RQ2: How effective are the social explanations in contextualizing \textsc{PaperPing}'s suggestions? (DG1)
    \item RQ3: How does \textsc{PaperPing} affect existing social practices in the channels? (DG2)
    \item RQ4: What impact do \textsc{PaperPing}'s suggestions have on channels' establishing common ground? (DG4)
\end{itemize}

\subsection{Study Participants and Procedure}
\subsubsection{Participants}
Via social media advertisement and word-of-mouth, we recruited a total of 28 participants from 18 channels (Table \ref{tab: participants_info}) to participate in interviews for our deployment study, ranging from junior PhD students to professors across various domains (e.g., AI, HCI, NLP, Programming Languages). The 
channels included group research labs, reading groups, project teams, and course channels, with sizes ranging from 2 to over 200 members each. These channels had varying levels of paper-sharing activities prior to installing the bot, ranging from 0 to over 5 papers per week. We paid participants who participated in both the interview and offline evaluation (N=24) \$50 Tango gift cards and those who participated in only the offline evaluation (N=4) \$25 Tango gift cards. 

Eighteen of our participants were interviewed before launching the tool in their channel. No participants were required to interact with \textsc{PaperPing} during the length of the study. 
 We additionally recruited 10 participants for interviews after the tool had been installed in their channel for a period of time by sending recruitment messages to the channel. These participants included those who indicated that they had not thus far used \textsc{PaperPing} much.

\subsubsection{Installing \textsc{PaperPing} in channels}
We invited participants who volunteered to be their channel's contact person to an onboarding session, where we explained \textsc{PaperPing}'s key features and guided them through installation and setup. To set up, participants selected a recommendation frequency suited to their channel's norms (e.g., once a week). We also invited participants to share several papers in the channels to seed \textsc{PaperPing} 
if the channel did not have much of a conversation history. Upon installation, \textsc{PaperPing} introduced itself to the group with a brief message, highlighting core features such as link sharing and emoji reactions. During the study, we sent check-in surveys every few weeks to all paid participants. These surveys captured recent memorable positive and negative experiences, which we used in post-study interviews to help participants recall their experiences.

\subsubsection{Exit interview and questionnaire}
After using \textsc{PaperPing} for at least 3 months, we invited our 28 participants to reflect on their experience in an exit interview lasting 30--75 minutes. During the session, they first completed the offline evaluation task (below), after which they discussed their experience in a semi-structured interview focusing on their perceptions of PaperPing's alignment with their individual and group preferences and its impact on group dynamics. 

Finally, they completed a questionnaire containing Likert scale questions about their overall satisfaction in and the usability of \textsc{PaperPing}. We developed survey questions by reviewing prior work on the evaluation of conversational AI tools~\cite{avulaEmbeddingSearchConversational2019}.

\subsubsection{Offline evaluation task}

We designed an offline comparative experiment to examine the effect of different variations of messages on participants' experiences. 
In the task, we presented four variations of explanations for papers (see Figure~\ref{fig:condition_figures}). Specifically, Condition 4 (LLM synthesis of social grounding information and paper summary) shows \textsc{PaperPing}'s messages posted in the  Slack channel. Condition 1 (TLDR), Condition 2 (social grounding information presented in template form), Condition 3 (social grounding information and paper summary information presented in template form) each explore a different component of a generated message and their effects on users' perception.
Each participant was asked to evaluate five papers that were randomly selected from the recommendations that they had received in the channel. Each paper was shown in four variations (Figure ~\ref{fig:condition_figures}), presented in random order. For each variation, participants rated Likert scale questions, such as whether the explanation was concise, interesting, or helped them understand the relevance to the group. After seeing all four variations, they selected the variations that best suited their own and group preferences, respectively.

\subsection{Data Collection and Analysis}
\subsubsection{ Qualitative coding for exit interviews}
We took detailed memos during the interview process and discussed emerging themes in regular project meetings. We used Dovetail\footnote{https://dovetail.com/} to transcribe the interviews, which produced transcriptions that were linked to audio and video recordings. During the coding process, we referred back to the original video recordings when transcripts were ambiguous or inaccurate.  We analyzed the data following the guidance of reflective thematic analysis~\cite{braunReflectingReflexiveThematic2019}, wherein we first developed codes based on the research questions and emerging themes in the memos and added new codes as they emerged in the analysis process. The final codebook contained 35 codes focused on: alignment of recommendations with participants' preferences, the effectiveness of the social signals, and \textsc{PaperPing}'s impact on social dynamics; examples include: \textit{relevance to individual preference}, \textit{conciseness of explanations}, and \textit{individual willingness to engage with \textsc{PaperPing}}.

\subsubsection{Statistical analysis of offline evaluation task ratings}
\textsc{PaperPing} made 832 recommendations in 18 channels during the deployment study. We randomly selected 136 of these papers and collected 540 observations from 28 participants in the offline evaluation task. Participants' ratings for this task were treated as ordinal variables and analyzed using a mixed ordinal logistic regression, where variations were fixed effects and participants were treated as random effects. We calculated the pairwise differences between the four conditions and applied Tukey correction. We also applied a Holm correction for testing multiple hypotheses.

\subsubsection{Usage logs}
To gain insight into how people interact with \textsc{PaperPing} in real-world settings, we collected and analyzed paper-related activities (e.g., reactions and comments to shared papers) within each channel, until all participants in the channel had completed the offline evaluation tasks and exit interviews. Notably, while we captured activities around papers shared via common academic platforms (e.g., ACM Digital Library, Semantic Scholar, arXiv), we were unable to track papers shared through social media links (e.g., X, LinkedIn posts).

\begin{figure*}[h]
    \centering
    \includegraphics[width=.85\textwidth]{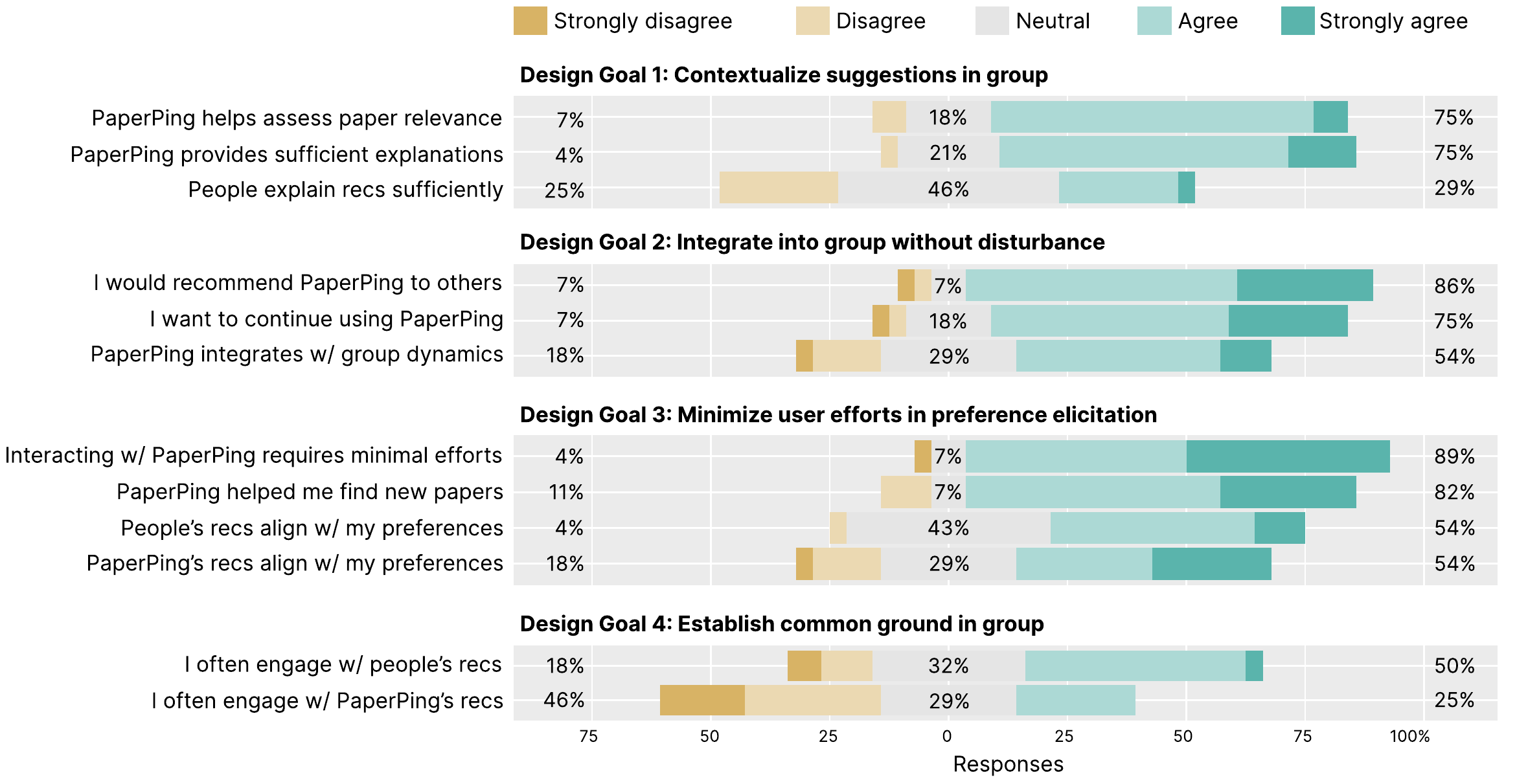}
    \caption{\textbf{Results from post-study questionnaire.} Responses are grouped based on the four design goals. We present the original questions posed by the questionnaire in the appendix.}
    \Description{Bar chart displaying user feedback on \textsc{PaperPing} across four design goals: 1) Contextualize suggestions in group; 2) Integrate into group without disturbance; 3) Minimize user efforts in preference elicitation; 4) Establish common ground in group. Responses are categorized as Strongly Disagree, Disagree, Neutral, Agree, and Strongly Agree. Key findings are discussed in paper.
}
    \label{fig:questionnaire}
\end{figure*}

\subsection{Results}

\subsubsection{\textbf{\textsc{PaperPing} effectively learns individual and group preferences with minimal user effort (DG1, DG3)}}

\paragraph{\textsc{PaperPing} recommends papers that suit individual and group preferences and that directly benefit ongoing research}

\textsc{PaperPing}'s messages were considered relevant and aligned well with researchers' personal and group preferences. In the offline evaluation, 59.6\% of the 136 reviewed recommendations were marked as relevant to participants' individual preferences, and 75.7\% were marked as relevant to groups' preferences (Figure~\ref{fig:questionnaire}). 
Moreover, 82\% of participants agreed that \textsc{PaperPing} helped them discover papers they would not have found otherwise. Over half (54\%) of the participants thought \textsc{PaperPing}’s suggestions aligned with their interests, which is comparable to those who consider recommendations from other channel members to be relevant. 
In fact, \textsc{PaperPing}'s suggestions were able to directly benefit researchers' ongoing research.
Participants considered benchmark papers suggested by \textsc{PaperPing} in their research (P8, P11): ``\textit{We definitely use the papers that are suggested.}'' Several participants (P5, P15, P16) appreciated \textsc{PaperPing}'s ability to surface relevant papers in fast-moving fields. Compared to general-purpose literature discovery tools, \textsc{PaperPing} was considered more helpful since it is ``\textit{seeded}'' and ``\textit{ha[s] a good sense of the topics and research communities that I want to read papers from}'' (P3). P11 echoed: ``\textit{With \textsc{PaperPing} we can focus our time and energy on our own paper.}''

\paragraph{Indicating one's preference to \textsc{PaperPing} involves minimum effort}
While \textsc{PaperPing} suggests relevant content that brings practical benefits, researchers agreed that indicating their interests to \textsc{PaperPing} was a natural process without being effortful. In the post-study questionnaire, 88\% of participants agreed that using \textsc{PaperPing} required little effort. Sharing paper links in the channel was the most frequently used method to communicate interests to \textsc{PaperPing}. 
Participants appreciated that these feedback mechanisms fit into their existing workflows. P3 described reading and sharing papers in channels to be a ``\textit{natural part of my day-to-day interaction of being on Slack.}'' P4 also thought that reacting to papers in group chats is a natural interaction: ``\textit{it feels social...other people also are interested in this, and they also react to this.}'' P20 described group members' emoji reactions and comments as a form of ``\textit{collective annotation}.'' 

\paragraph{Limitations of existing feedback mechanisms}
Despite promising results, we observed several limitations of \textsc{PaperPing}'s existing feedback mechanisms. 
In less active groups, \textsc{PaperPing} struggled to gather sufficient signals, leading to less accurate recommendations. 
P10 mentioned: ``\textit{there are definitely more interests that we have, but we just didn't put in [these in] the channel, so \textsc{PaperPing} wasn't able to pick up the signal.}'' P5 and P13 both wished their group members had reacted to papers more frequently to better inform \textsc{PaperPing} about their groups' interests. Moreover, diverse group norms about using emojis could introduce ambiguity in the interpretation of the feedback. For example, P8 noted that in their channel, ``\textit{the reaction was more for acknowledgment as opposed to indicating that the paper is very useful.}'' Further, although \textsc{PaperPing} considers all papers shared in the past in order to have more data to be able to use, it struggled to reflect interest shifts in a timely manner: ``\textit{It's hard to branch out into super new papers...it feels like the types of papers that [it] recommend[s] don't change too much over time}'' (P1). 

For groups with diverse interests, the tension between individual and group interests also poses challenges. P4 observed: ``\textit{even paper shared by my lab mates I don't relate to most of the time.}'' As a result, \textsc{PaperPing} sometimes ``\textit{overfits}'' to certain channel members' interests (P5). The misalignment could further turn into barriers for users with unique interests to share papers. P17 felt uncomfortable sharing papers that diverged from the group's primary focus, which represented the interests of only active members in the channel.

\paragraph{Despite recognizing \textsc{PaperPing}'s usefulness, participants did not see it as a replacement for paper recommendations from other human channel members.} 
 As P3 describes: ``\textit{\textsc{PaperPing} is not a replacement. It's an addition to the practice of people sharing with other people.}'' Participants noted that \textsc{PaperPing} did complement human recommendations in its timeliness, broad coverage, and ability to tie back to previous discussions. P8 shared: ``\textit{there are some papers that [\textsc{PaperPing}] suggested that I think people just missed or they didn't suggest it in the group. [\textsc{PaperPing}] helps to have a better picture of the ongoing work related to the project.}'' P4 observed that \textsc{PaperPing} offered more diverse and more timely recommendations than they would typically receive from human group members. 
However, participants continue to place more trust in human recommendations and find they want to  verify recommendations from \textsc{PaperPing}. For example, P8's comparison of papers from \textsc{PaperPing}, arXiv, and other human members further illustrated this sentiment: ``\textit{\textsc{PaperPing} is more helpful than browsing arXiv since the topics are more focused, but I still feel like I have to briefly judge whether I wanna actually read it or not. Whereas if a person who is familiar with our project suggested it, I would definitely read it without having to consider it again.}''

\begin{figure*}[h]
    \centering
    \includegraphics[width=.8\textwidth]{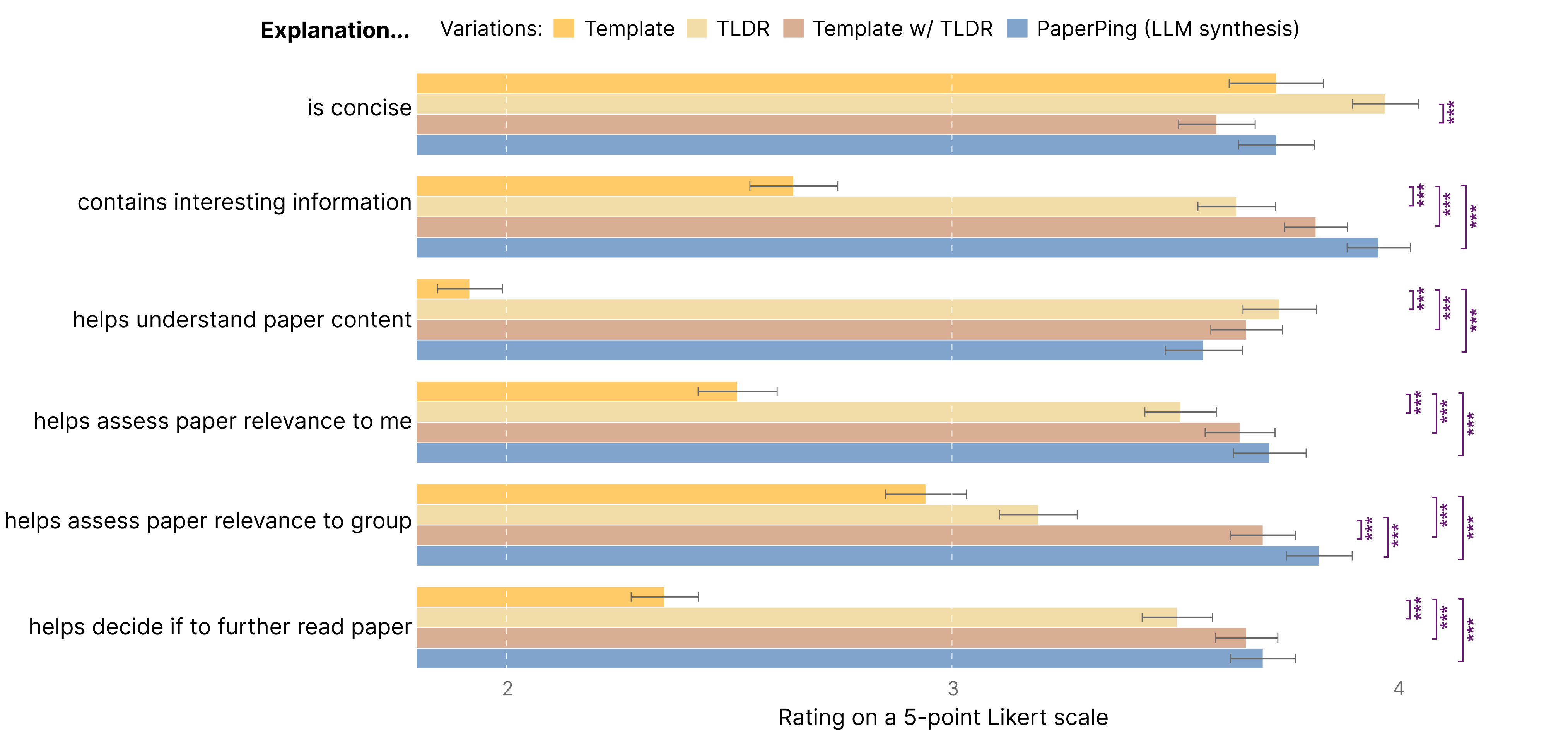}
    \caption{\textbf{Results of our offline evaluation. }Error bars show standard errors of the mean (SEM). 
    The ``template'' messages (Condition 2) were consistently rated lower than \textsc{PaperPing}’s LLM synthesis (Condition 4) and Template+TLDR (Condition 3) across all questions ($p$<0.001), except for conciseness. The ``template'' messages (Condition 2) were also rated lower than TLDR (Condition 1) across all questions ($p$<0.001), except for conciseness and on ``explanation helps assess paper relevance to group.''. Additionally, explanations with social signals (\textsc{PaperPing}’s LLM synthesis (Condition 4) and Template+TLDR (Condition 3)) were rated significantly more helpful on ``explanation helps assess paper relevance to group'' compared to the TLDRs, which only summarized paper content ($p$<0.001 for both comparisons). We observed no  statistical significance in other comparisons. 
    }
    \Description{Bar chart comparing user ratings from the offline evaluation results for different explanation variations: Template, TLDR, Template with TLDR, and \textsc{PaperPing} with LLM synthesis. The explanations are evaluated on several aspects: conciseness, containing interesting information, helping understand paper content, helping assess paper relevance to the individual and the group, and helping decide whether to read the paper further.
}
    \label{fig:stats}
\end{figure*}

\subsubsection{\textbf{Social explanations contextualize suggestions in group context (DG1)}}

\paragraph{Social signals in explanations help contextualize the recommendation in social context}
To adapt AI suggestions to groups' social contexts, \textsc{PaperPing} explained each paper recommendation bysummarizing paper content and its relevance to the group. These explanations helped participants assess paper relevance (75\% agree) and were perceived as sufficient by most participants (75\% agree). In contrast, only 29\% of participants' explanations provided by other channel members were sufficient (Figure~\ref{fig:questionnaire}). In the offline evaluation, explanations with social signals (\textsc{PaperPing}'s LLM synthesis and Template+TLDR) were rated as being significantly more helpful for conveying relevance to the group compared to the TLDRs, which only summarized paper content ($p<0.001$ for both comparisons; see Figure~\ref{fig:stats}).

In the interviews, participants shared how social signals helped them determine a paper’s relevance. 
For example, P10 and P11 mentioned inferring paper relevance by using the importance of the linked paper (i.e., similar or cited by the recommended paper) and the presence of shared authors highlighted in the explanation. 
Others also mentioned using the channel engagement of the linked paper highlighted in the explanation to determine paper importance: ``\textit{three thumbs up versus one neutral reaction makes a difference}'' (P4). 
P17 noted that social signals could be more helpful than paper content summaries in determining relevance: ``\textit{I only need to glance at the people who are mentioned; I don't even need to go through the summary of the paper to know how relevant this paper is to me.}'' This can be especially helpful for junior scholars like P21, who have less expertise on the topic compared to other channel members: ``\textit{I don't have as much background on the main topic of the channel, so social signals were helpful for me in understanding why it'd be relevant for me to read.}''
As P21 summarized: ``\textit{social signals provided a more contextual way of understanding how the recommended paper related to a past paper.}''

Social signals also foster group members' trust in \textsc{PaperPing}'s suggestions. P8 offered that social signals provided transparency into \textsc{PaperPing}'s actions: ``\textit{the fact that there are correct links to who reacted to or who suggested what helps you to trust that...the information is being considered in an accurate or relevant way.}'' 
The association with specific group members highlighted in the explanations made P17 more receptive to \textsc{PaperPing}'s suggestions not immediately relevant to them: ``\textit{if \textsc{PaperPing} mentioned a person who's not in my small group or who I don't work with, I know this paper is not relevant to me but is relevant to the larger group.}''

\paragraph{\textsc{PaperPing}'s LLM synthesis is informative and natural but could lead to unreliable interpretation}
Participants thought \textsc{PaperPing}'s LLM-synthesized explanations could introduce paper content in personalized and contextualized ways and surface connections that might otherwise be missed. P3 appreciated the LLM explanations that highlighted the novelty of the recommended paper beyond its findings. P5 and P13 valued the explanations and connections with their interests that they had not previously considered: ``\textit{That made me decide that I could take a look at the recommended paper}'' (P13).
However, some participants found LLM explanations to be too speculative or ambiguous. P10 noted that LLM synthesis could overly compress information, making the explanations unclear.
P12 also found LLM explanations to be too generic: ``\textit{It highlighted the similarities between two papers but can be too broad, like `boosting computer visual models.' That's what every single paper is about.}'' Indeed, the usefulness of social signals relies on the prerequisite of having a clear summary of the paper's content. This is evident in our offline evaluation, where the template condition (Condition 2) was consistently rated lower than both PaperPing’s LLM synthesis (Condition 4) and Template with TLDR (Condition 3) across all evaluation questions except conciseness ($p<0.001$ for all comparisons; see Figure~\ref{fig:stats}). It was also rated lower than TLDR (Condition 1) on all questions ($p<0.001$) but conciseness and helpfulness in assessing relevance to the group. For many researchers, a clear summary of the paper is considered to be a ``\textit{more reliable source of information}'' (P7) compared to inferred social signals (P1, P2, P7, P15).

Participants also noted that \textsc{PaperPing}'s awareness of individual and group context could still be insufficient. For example, P10 found that explanations occasionally highlighted connections that were misaligned with their interests: 
``\textit{it has its own understanding of why and sometimes that leads to inaccuracy or irrelevance.}'' 
P17 noted that \textsc{PaperPing} could miss group context cues and offer redundant information: ``\textit{Everyone in the group already knows [popular machine learning model] and has been following the different versions. It's just like announcing a new iPhone; you don't really need to introduce the iPhone 14, iPhone 13 before that. You can just say, `Here's the new iPhone 15.'}'' 
Participants P6 and P18 observed that the LLM's interpretation of emoji reactions in explanations could be awkward and inaccurate: ``\textit{It feels like it's trying to derive the conclusions when not having enough data points.}'' (P1); they would prefer to see the exact emoji to better understand the nuances of the reactions, especially given the diverse norms of emoji usage across groups. 
Due to these uncertainties in LLM interpretations, participants, including P10, preferred the TLDR+template explanations (Condition 3) because ``\textit{It felt a little bit more objective.}'' P14 also preferred forming their own interpretations after reading a paper instead of relying on \textsc{PaperPing}'s interpretation.

\begin{figure*}[h]
    \centering
    \includegraphics[width=.95\textwidth]{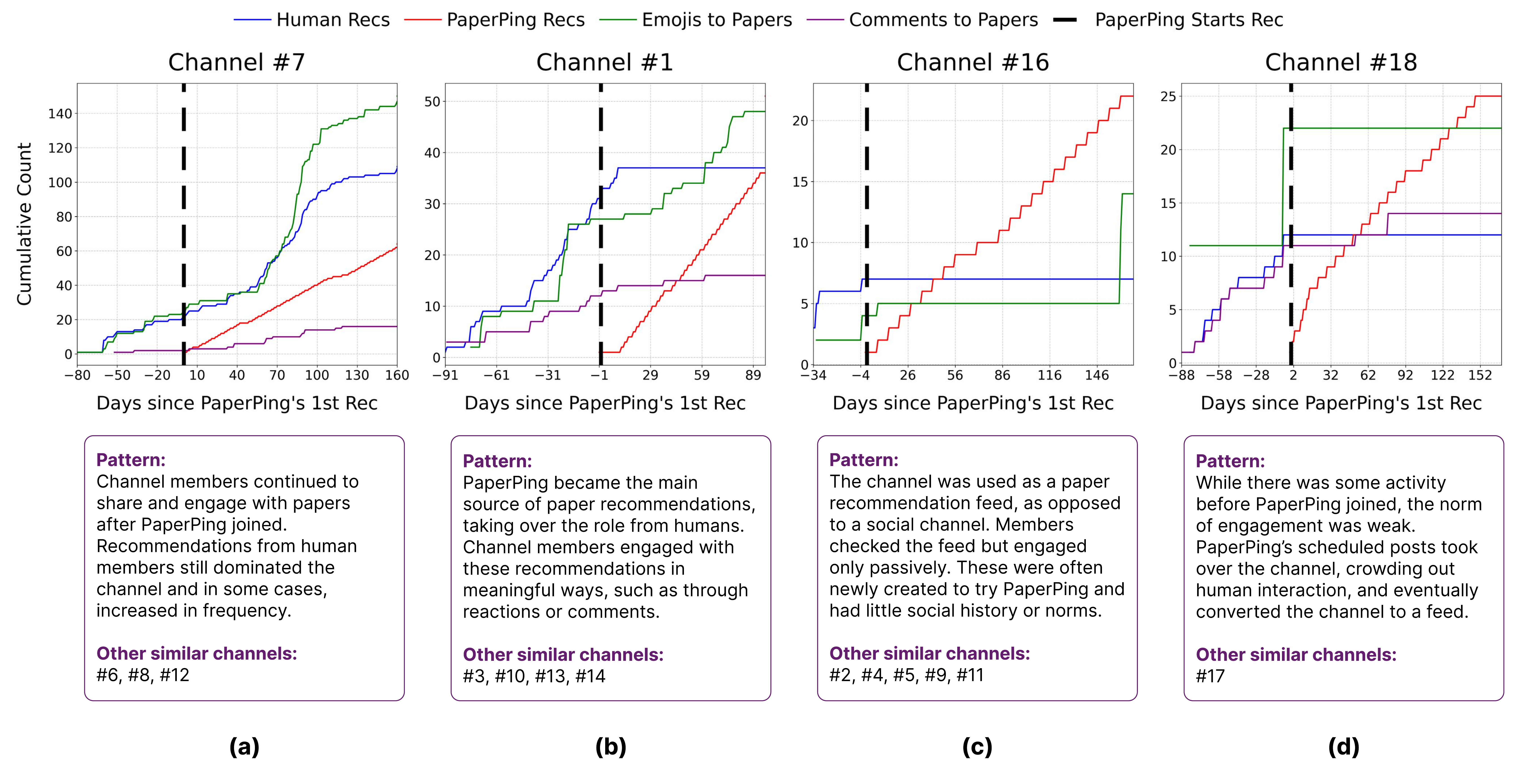}
    \caption{\textbf{Representitive patterns of usage of \textsc{PaperPing} during the deployment study.} The y-axis represents the cumulative counts of usage in each channel, and the x-axis represents the days since \textsc{PaperPing}'s first recommendation. Notably, only activities around papers shared via common academic platforms (e.g., ACM Digital Library, Semantic Scholar) are captured. We were unable to track papers shared through social media links (e.g., X, LinkedIn posts).}
    \Description{Four line charts showing cumulative paper-related engagements in representative Slack channels over time. Each graph shows four lines, representing the cumulative number of human recommendations (blue), PaperPing recommendations (red), emoji reactions (green), and comments (purple). A vertical dashed line at day 0 indicates the time when PaperPing was installed. Channels show pre-existing paper-sharing activity before PaperPing installation and varying rates of growth in different metrics afterward}
    \label{fig:usage}
\end{figure*}

\paragraph{Challenges involved in translating social explanations to personal relevance}

Though social signals can help contextualize a paper's relevance to the group, participants P1 and P19 noted the challenge of translating them to their personal interests. As P19 explained: ``\textit{If I decided to open up that paper, those reasons are all related to myself.}'' Seeing \textsc{PaperPing}'s social explanations related to other group members could help them infer the paper's relevance to themselves, but P1 considers it challenging to do the translation: ``\textit{We have very diverse interests, and people have different standards for liking. I don't know how they relate to my own standards.}'' 
Indeed, the utility of social signals in explanations varies across individual members in groups. P17 and P22 found references to past papers confusing if they had joined the channel later or forgotten the shared context: 
``\textit{To me that shared context was lost}'' (P17). P22 highlighted that ``\textit{shared context from working in the same team towards the same goal}'' was essential for them to find information about relevance to specific members to be helpful: ``\textit{if I had not been there in person or if I didn't know these people, then I don't know how much it would help.}''

\subsubsection{\textbf{For most groups, \textsc{PaperPing} does not disrupt existing social practices (DG2)}}

\textsc{PaperPing} 
generally fits in well with a group's dynamics, as more than half of the participants believe (54\% strongly agree or agree).
86\% of participants indicated they would recommend \textsc{PaperPing} to others, and 75\% expressed a desire to continue using \textsc{PaperPing} in their channels. In fact, 11 out of the 18 participating channels are still voluntarily using \textsc{PaperPing} as of this writing, more than six months after the conclusion of our study. As P4 described: ``\textit{\textsc{PaperPing} feels natural to me. It doesn't disrupt the natural flow of the conversation in the channel.}'' P1 described \textsc{PaperPing} as ``\textit{an invited robot guest in the group's space, adopting to groups' norms rather than trying to make the group adapt to it.}'' They elaborated: ``\textit{It doesn't really feel out of place. It recommends things in a message and links back to previous threads just like any other member.}'' P23 echoed the sentiment: ``\textit{\textsc{PaperPing}'s behavior didn't interfere with anything. It didn't disrupt the natural conversation.}'' \rw{Indeed, the usage logs of Channel \#7 and Channel \#1, which P23, P1, and P4 belong to, respectively, along with other channels, indicate that members continued to engage with both papers shared by their peers and those recommended by \textsc{PaperPing} (Figure~\ref{fig:usage}(a) and (b)). Notably, in Channel \#7, recommendations from human members remained the dominant source of paper sharing, whereas in Channel \#1, \textsc{PaperPing} became the primary source of paper recommendations.} 

P4 appreciated the pace of posting: ``\textit{The frequency is nice. I didn't feel like I was annoyed by the bot or anything.}'' P8 and P11 noted that \textsc{PaperPing}'s contextualized, natural language explanations made it feel more human. Participants including P16 and P21 also liked the concise and stylized summary with key information bolded: ``\textit{I like bolded text in the explanations. [It is] especially helpful for Slack messages, allowing me to skim it first and then...decide if you want to read it.}'' (P21)
However, other participants found \textsc{PaperPing}'s explanations could be verbose: ``\textit{If there's a huge chunk of text in Slack, I tend to glance at the first sentence or maybe word. If it's not important, then I just dismiss it.}'' (P15). The chunk of text could potentially bury human conversations if not posted at an ideal time (P23). \rw{This highlights a limitation of \textsc{PaperPing} that emerged in the field study, which we observed in two participating channels (Figure~\ref{fig:usage}(d)). While channel members could set the posting frequency, \textsc{PaperPing}’s scheduled messages sometimes overwhelmed users, inadvertently crowding out human interactions, particularly when existing social norms  around engagement are fragile. We discuss this limitation and its design implications in the Discussion Section. }

\subsubsection{\textbf{\textsc{PaperPing} fosters common ground within groups (DG4)}}

\paragraph{Social signals help facilitate the establishment of common ground}
\textsc{PaperPing} not only suggests useful and relevant information without disrupting channels but also plays a social-facilitator role by exposing group members to each other's interests and fostering common ground within groups. Participants noted that social signals, such as tagging relevant members and referencing prior discussions and engagements, could lead to potential follow-up interactions and foster group cohesion.
As P14 reflected, 
``\textit{The social signal is the strongest aspect because it shows relevance to me or to my group, which helps me understand my group pattern to some extent.}''  
Participants appreciated how \textsc{PaperPing}’s social signals facilitated group members learning more about each others' interests. As P3 shared, `\textit{The recommendations and the social signals helped me realize my group members were interested in this type of research.}'' 
Integrated into existing social spaces and practices, \textsc{PaperPing} encourages researchers to develop further interactions among each other. 
As P14 mentioned, ``\textit{\textsc{PaperPing} is a part of a chat group with other members. I get to understand the interests of other people. It's a part of a practice that I already have in real life, and it has a lot of potential to assist and support in connecting better with [each] other.}''
P10, a member in a large research lab channel, also sees \textsc{PaperPing}'s explanations as opportunities to develop long-term collaborations: ``\textit{Just having it [\textit{\textsc{PaperPing}] in the group is helpful for me to know what other people's interest is. Even if the interest is slightly different, seeing some new things from related fields is helpful for me to learn what other people are working on, common things that we work on, and things that we could be chatting about.}}''

\paragraph{\textsc{PaperPing} reinvigorates activity in inactive channels but does not necessarily establish new norms}
Several participating channels were lab groups that wanted to encourage paper-sharing but have not to date been successful. In these 
inactive channels where there was little conversation among members, \textsc{PaperPing} helped revive activity by posting recommendations and tagging members to prompt engagement. Participants felt that \textsc{PaperPing} ``\textit{Keep[s] the channel alive}'' (P3, P14). As P14 described: ``\textit{Whenever there was a new message, I would look at it.}'' P10 observed that \textsc{PaperPing} prompted more contributions from channel members: ``\textit{The channel hasn't been active at all for half a year or more. \textsc{PaperPing} definitely prompts some other people to post. At least I'm seeing two or three other people who have posted on this channel.}'' Indeed, P8 thought that \textsc{PaperPing} served as a reminder for other human members to engage: ``\textit{It reminded people to send paper recommendations. We have always been encouraged to do that, but people tend[ed] to forget previously. Seeing \textsc{PaperPing}'s recommendation every two days in the channel reminded people.}'' \rw{The usage log of Channel \#7, to which P8 belongs, shows that channel members' frequency of sharing and reacting to papers increased after \textsc{PaperPing} joined the channel} (Figure~\ref{fig:usage}(a)). P13 added that \textsc{PaperPing} incentivized them to contribute more: ``\textit{It made me feel my reactions or activities in the channel are being seen and are going somewhere, which gets me excited to contribute. It will be captured by \textsc{PaperPing}, and sometime later it is going to give me a reward.}'' 
Participants also noted that tagging specific channel members is an effective way to keep people engaged;
P5 usually skims through \textsc{PaperPing}'s posts, but ``\textit{If my name is tagged, I would read a little more into it.}''

\paragraph{However, participants also noted that \textsc{PaperPing} did not establish new norms, such as having discussions about papers in groups, that were not previously established practices.} Only 25\% of the participants self-reported often engaging with \textsc{PaperPing}’s recommendations, compared to 50\% of the participants who self-reported often engaging with other human recommendations (Figure~\ref{fig:questionnaire}).
P15 observed that although \textsc{PaperPing} was helpful in suggesting relevant papers, this did not lead to more paper discussions in the channel.  
In fact, several other channels also show a similar engagement pattern with Channel \#4, to which P15 belongs  (see Figure~\ref{fig:usage}(c)). These channels often have little social history or established norms, with several created right before the study or inactive for an extended period of time. Instead of fostering a social space, these groups used \textsc{PaperPing} to create a feed of paper recommendations, which members checked regularly but only engaged with passively.   
The lack of extrinsic incentives could be one reason for less engagement with \textsc{PaperPing}'s recommendation. Although we disclosed \textsc{PaperPing}'s feedback mechanism to the whole group upon installation, P8 noted that some group members could be unaware that their engagement could inform future recommendations and thus were not incentivized to engage. Even when aware of the mechanism, participants could still be hesitant to engage because of the lack of potential follow-up interactions with \textsc{PaperPing}: ``\textit{I know \textsc{PaperPing} won't respond.}'' 
Participants also agreed that respected human members were more effective at encouraging engagement than \textsc{PaperPing} alone. P4 observed that their channel became more receptive to \textsc{PaperPing}’s recommendations and had more engagement after their advisor reacted to one of the posts.

\section{Discussion}

Our evaluations of \textsc{PaperPing} showed that an agent that is augmented with Social-RAG can deliver contextually relevant and socially aligned messages. 
\textsc{PaperPing}'s strategies to convey relevance to groups, such as referencing prior discussions and highlighting group members' interests, effectively contextualized the message to suit  group social norms (DG1). 
Its ability to learn interests based on past paper sharing history and input ongoing feedback via emoji reactions, comments, and shared links let users provide continuous feedback with minimal effort (DG3). 
Users' interactions with \textsc{PaperPing} (e.g., receiving messages in group chats) naturally suited their existing social practices (DG2) and helped them understand other group members' interests, occasionally even reviving group activity (DG4). 

Nonetheless, participants expressed interest in even more nuanced levels of social awareness, such as improvements in distinguishing between group and individual preferences (DG1), tailoring the verbosity and tone of messages based on group norms (DG2), and enabling more group-specific feedback mechanisms (DG3).
Based on our findings, we discuss the implications for future systems that plan to support socially nuanced text generation.

\subsection{Bootstrapping Social Grounding with Past and Ongoing Group Interactions}
The use of past and ongoing group interactions as a basis for social grounding is a core advantage of Social-RAG. In contrast to explicit preference elicitation, which requires significant user effort~\cite{nguyenDynamicElicitationUser2017} and risks disrupting group dynamics~\cite{agarwalConversationalAgentsFacilitate2024}, Social-RAG infers user preferences from past and ongoing group social interactions without requiring additional user labor. Leveraging collected user preferences, or \emph{social facts}, Social-RAG uses an LLM to generate contextually relevant messages. For this, we developed techniques to curate and maintain a social knowledge base from past and ongoing group interactions. For example, the PaperPing system preserves the connection between extracted items (e.g., shared papers) and their social contexts (e.g., reactions and comments from others) when indexing group conversation history into the social knowledge base. These preserved connections enable further social inference in the Social-RAG workflow. We also proposed complementing the off-the-shelf retrieval methods typically used in traditional RAG systems (e.g., semantic embeddings~\cite{lewis2020retrieval}) with empirically validated heuristics to guide the retrieval and interpretation of social knowledge. This hybrid approach, which combines heuristics with off-the-shelf embedding methods, is also starting to show promise in recent RAG applications in other contexts that require more complex reasoning (e.g., UI linting~\cite{lu2024ai}).

\subsection{Costs and Benefits of Different Levels of Group Social Grounding}
Grounding systems in their social contexts has been highlighted as a major design requirement for both classic social computing systems~\cite{ackerman2000intellectual} and recent LLM-powered systems~\cite{vaithilingam2024imagining}. However, realizing and implementing social grounding effectively remains an open challenge~\cite{vaithilingam2024imagining}. We contribute to this ongoing discussion by reflecting on the different levels of social grounding of generated text in a group setting based on our experience of developing Social-RAG and implementing \textsc{PaperPing}. Different levels may require different amounts of resources and considerations and could potentially lead to different benefits.

\textbf{No social grounding.} At the lowest level, a one-size-fits-all system could apply uniform behavior across all groups, relying on generic templates or heuristics (e.g., generate template text that is likely acceptable across all settings). This level of grounding is resource friendly and predictable, but it lacks the ability to adapt to specific social contexts. 

\textbf{Category-level social grounding.} A system could pre-define  more detailed rule-based heuristics to generate content for groups based on their broad characteristics, informed by empirical observations. For example, in the deployment of \textsc{PaperPing}, we observed that groups working on active research projects would prefer exhaustively exploring closely related papers on certain topics, but interest-based reading groups would prefer exploring diverse sets of papers. However, it would be challenging for systems at this level to adapt to nuanced differences in similar categories. 

\textbf{Group-level social grounding.} Further, a system could tailor its output to group-specific preferences and norms. In Social-RAG and \textsc{PaperPing}, we demonstrate that past and ongoing conversations in a group could effectively inform AI systems about the group's implicit topical preferences. A system at this level could also consider adapting the cadence and tone of generated messages to make them less intrusive to groups.

\textbf{Individual-level grounding.} Finally, a system could target and tailor its messages to specific group members (e.g., via mentions). While individual-level grounding could increase engagement and encourage understanding among group members, a system at this level needs to carefully design the balance between individual preferences and group preferences as well as carefully navigate power dynamics to behave in a socially appropriate way.

To more concretely illustrate the above, we can position \textsc{PaperPing} 
(i.e., Condition 4 in the offline evaluation) within this framework: Selecting which paper to recommend is grounded at the group-level by pooling all previously shared papers as the input to a recommender system. This gives group members access to papers of both common and individual interest. The generated explanations were tailored at both the group and individual level, with relevant members being mentioned. This promoted engagement with \textsc{PaperPing} as well as awareness of specific member's interests to all group members.

\subsection{The Tension of Group vs. Individual Preferences}
As previously noted, a key design consideration for a socially grounded system is how to balance group- and individual-level personalization. Empirical observations from our 3 months deployment showed that within the same group, individuals might have different preferences for explanations relative to each other or even prefer different messages. 
For example, members who are more familiar with a recommended paper might prefer explanations that are more in-depth compared to other members.

One way to address this tension is by aligning the generation with an ``average group'' preference. 
Future systems could consider experimenting with different ways to aggregate individual preferences that optimize various goals.
Another option is to scaffold the explanation, e.g., starting with a more generic explanation tailored to all members and following with in-depth explanations tailored to individual members.
However, if such modifications are made, the potential risk of diminishing common ground among members must be carefully considered (DG4).

\subsection{The Impact of Proactive Agents on Group Social Dynamics}
Another key design goal of a socially grounded system is ensuring that AI agents enhance rather than crowd out desirable human-human interactions. In our deployment of \textsc{PaperPing}, we observed varied impacts on group dynamics. In some cases (Figure~\ref{fig:usage}(a) and (b)), \textsc{PaperPing} had a positive influence. Channel members became more active in sharing and reacting to papers or offloading paper-sharing to \textsc{PaperPing} while continuing to engage through reactions or comments. This suggests that AI agents can play a beneficial role in group spaces by taking on tedious tasks while preserving human engagement.
However, as seen in Figure~\ref{fig:usage}(c), the presence of \textsc{PaperPing} alone did not automatically foster new social norms (e.g., active paper discussion), particularly when human members primarily valued receiving recommendations over engaging in conversations. Moreover, in some cases (Figure~\ref{fig:usage}(d)), a mismatch between \textsc{PaperPing}’s scheduled posts and the natural pace of group interactions could overwhelm the channel, leading to disengagement and abandonment of the space.
Overall, these observed patterns point to a promising future of AI agents taking on pro-social roles in groups but highlight the importance of designing agents that can flexibly adapt their content and behavior (e.g., pacing and timing of posting) to different group dynamics.

\section{Limitations}
While our participants found \textsc{PaperPing} to be useful in real-world online chat groups, our field deployment also highlighted limitations of our current implementation of Social-RAG that can inform interesting future work. 
Specifically, observable social interactions in conversation history may not fully capture group dynamics or offline interactions, potentially leading to biased or inaccurate interpretations of preferences and norms. For example, in less active groups, \textsc{PaperPing} had fewer social signals to work with (e.g., shared papers, user reactions, and comments) and produced lower quality recommendations and explanations. Similarly, as group dynamics evolve over long periods of time, the relevance of past interactions may become stale or diminish even for active groups. Some participants reported finding it hard to shift to new topics, and some even created a new channel to indicate new interests to \textsc{PaperPing}.

\section{Future Work}

While this work focused on leveraging social signals to tailor the right information with the most relevant social context, future work could  consider additionally tailoring the tone of explanations, how many details to include, group posting cadence, etc. Similarly, the current implementation of \textsc{PaperPing} focuses on aligning with a group's topical preferences, but one could also extend the workflow to account for, say, social hierarchies. The challenge would be to explore relevant social signals to extract from the group's interaction history and find ways to effectively use them as part of the context for LLMs to generate socially appropriate messages.

Another area of future work would be to explore more flexible ways to adapt to a group's communication norms. For example, we currently hard-coded how the system interprets emoji feedback from users (e.g., a thumbs-up emoji indicates positive interest in the paper), a method informed by the norms observed in the formative interviews.
However, our deployment study showed that a few individuals or groups did not follow this assumption (e.g., using a thumbs-up emoji only as an acknowledgment). Future work could explore how agents could more flexibly interpret social signals based on observation of groups' past social interactions.
In sum, the flexibility to ground different aspects of a system at varying levels offers significant opportunities for creating more socially aware and contextually relevant AI agents. However, it also requires thoughtful design choices and a deep understanding of the social dynamics occurring within target groups.

As with any system designed to be embedded in social spaces, learning from interactions introduces privacy concerns. 
One way to mitigate these is to limit the scope of what the system is allowed to learn from. For example, we made the conscious decision to allow \textsc{PaperPing} to access user interactions only on messages where a research paper was shared. This is a trade-off between privacy concerns and how well our system can learn from interactions between members.
Future work could provide greater user control over which aspects of their interactions are used to inform AI-generated content, potentially through customizable privacy settings or explicit consent mechanisms. 

Further, \textsc{PaperPing} does not currently allow users to manually insert into or edit the collected social signals. Future work could explore ways to encourage and facilitate user engagement in less active groups as well as complement Social-RAG with direct elicitation techniques to support scenarios where the group needed to make more drastic changes to its preferences (e.g., when switching focus to a new project.)

Finally, exploring Social-RAG in other collaborative spaces and scenarios that are rich with social interaction histories is also an exciting future direction. For example, in Reddit communities, a misinformation correction bot empowered by Social-RAG could generate explanations that are best suited to each community's norms in order to convey trust. Specifically, when retrieving external documents, the system could learn about the types of sources (e.g., governmental vs. journalistic) that are often shared and trusted by the community as indicated by the sentiment of replies; when generating posts, the system could learn from successful debunking strategies that were used in the past by members. Another example of a Social-RAG system could be a meeting summary tool that can generate summaries and tag the most relevant team members for awareness. The tool could analyze past documents to understand team members' interests based on signals, such as contribution history and viewing frequency, or analyze meeting information, such as speaking frequency in discussions of different topics or speaker roles.

\section{Conclusion}
This work introduced a novel LLM workflow, Social-RAG, that enables AI agents to ground their generation in the social context of group interactions. By retrieving and leveraging ``social facts'' from past and ongoing group interactions, Social-RAG enables AI agents to dynamically adapt their outputs to group norms and preferences with minimum user effort. We instantiated this workflow in \textsc{PaperPing}, an AI agent that recommends scholarly papers and generates contextually tailored explanations in research group chats. Our 3-month deployments of \textsc{PaperPing} in 18 Slack channels demonstrated Social-RAG's ability to adapt to diverse group dynamics, fostering trust and common ground while minimizing disruption to existing social practices. Overall, our work demonstrates Social-RAG's potential to address the social-technical gap in designing adaptive and socially aware AI systems by introducing strategies to retrieve implicit group dynamics and generate contextually aligned content. Social-RAG moves us closer to a future with a more natural and effective integration of AI agents in collaborative environments.

\begin{acks}
We would like to thank the study participants for their valuable insights and the anonymous reviewers for their helpful feedback. We would also like to thank Chuqiao Sun and Yangtian Yan for their help in refining the figures in this paper. The work is supported by grants from the Allen Institute for Artificial Intelligence (Ai2) and Microsoft's New Future of Work Initiative.
\end{acks}

\bibliographystyle{ACM-Reference-Format}
\bibliography{reference}

\appendix

\section{Prompts designed for PaperPing}
To anonymize, social context information in the following examples is hypothetical. The paper abstracts in the example prompts have been shortened for readability in this appendix. The actual prompts contained the full abstracts. 

Prompt 1 in Fig.~\ref{fig:prompts} (\textit{before} adaption):
\begin{Verbatim}[breaklines,frame=single,numbers=left,xleftmargin=5mm]
You are a helpful assistant for paper summarization.
The paper has the following details:
* Abstract: [ABSTRACT_OF_RECOMMENDED_PAPER]
* Authors: [AUTHORS_OF_RECOMMENDED_PAPER]
In your summarization, you must
* Congratulate the following authors: [AUTHOR_NAME_TO_CONGRATULATE].
* Mention the following authors: [AUTHOR_NAME_TO_HIGHLIGHT].
* Mention [AUTHOR_AFFILIATION_TO_HIGHLIGHT].
* Mention the paper's conference or journal: [VENUE_TO_HIGHLIGHT].
* Keep your output less than [CHARACTER_LIMIT] characters.
* Be informative.
* If applicable, be specific about the numbers in the abstract that may refer to the proposed method's performance.
\end{Verbatim}

\noindent Example Prompt 1 (\textit{after} adaption):
\begin{Verbatim}[breaklines,frame=single,numbers=left,xleftmargin=5mm]
You are a helpful assistant for paper summarization.
The paper has the following details:
* Abstract: {By simply composing prompts, developers can prototype novel generative applications...[remainder of the abstract omitted for readability]}
* Authors: Tae Soo Kim, Yoonjoo Lee, Jamin Shin, Young-Ho Kim, Juho Kim
In your summarization, you must
* Congratulate the following authors: Tae Soo Kim.  
* Mention the following authors: Juho Kim.
* Keep your output less than 350 characters.
* Be informative.
* If applicable, be specific about the numbers in the abstract that may refer to the proposed method's performance.
\end{Verbatim}

\noindent Output of example Prompt 1: \texttt{Congratulations to @tsk and team, including Juho Kim, on developing EvalLM! This system aids in refining LLM prompts by evaluating outputs against user-defined criteria, showing a 59\% reduction in needed revisions. It enables examining twice as many outputs, enhancing prompt development efficiency.}

\vspace{1em}

\noindent Prompt 2 in Fig.~\ref{fig:prompts} (\textit{before} adaption):
\begin{Verbatim} [breaklines,frame=single,numbers=left,xleftmargin=5mm]
You are a helpful assistant.
Title: [TITLE_OF_RECOMMENDED_PAPER]
Abstract: [ABSTRACT_OF_RECOMMENDED_PAPER]
Title: [TITLE_OF_RELEVANT_PAPER]
Abstract: [ABSTRACT_OF_PREVIOUS_PAPER]
The first paragraph of your answer should explain and specify the relationship between [TITLE_OF_RECOMMENDED_PAPER] and [TITLE_OF_RELEVANT_PAPER]. Be informative.
* In this first paragraph of your answer, you must explain and specify how [TITLE_OF_RECOMMENDED_PAPER] is related to [TITLE_OF_RELEVANT_PAPER] in one short sentence.
* In this first paragraph of your answer, you must start with "This paper might be related to [TITLE_OF_RELEVANT_PAPER] because".
* In this first paragraph of your answer, you must specify how [TITLE_OF_RECOMMENDED_PAPER] cites [TITLE_OF_RELEVANT_PAPER] in one short sentence. The content from [TITLE_OF_RECOMMENDED_PAPER] that cites [TITLE_OF_RELEVANT_PAPER]: [CITATION_CONTEXT].
* In this first paragraph of your answer, you must mention the following shared authors of the two papers: [SHARED_AUTHOR_NAMES].
* The first paragraph should have no more than [CHARACTER_LIMIT_P1] characters.
The message from [USERNAME] who shared [TITLE_OF_RELEVANT_PAPER]: [MESSAGE]
People's comments about [TITLE_OF_RELEVANT_PAPER]: [COMMENTS]
People's reactions about [TITLE_OF_RELEVANT_PAPER]: [EMOJIS]
The second paragraph of your answer should specify what people think about [TITLE_OF_RELEVANT_PAPER] and who these people are. Be informative.
* In this second paragraph of your answer, you must start with "thoughts about [TITLE_OF_RELEVANT_PAPER]". Note that user A 'cc' user B means that A thought [TITLE_OF_RELEVANT_PAPER] is related to B's research, projects, or interests. Thoughts would not be negative.
* In this second paragraph of your answer, you must appreciate that [USERNAME] shared [TITLE_OF_RELEVANT_PAPER].
* In this second paragraph of your answer, you must NOT add in-line citations and citation numbers.
* The second paragraph should have no more than [CHARACTER_LIMIT_P2] characters.
Your answer should replace [TITLE_OF_RECOMMENDED_PAPER] with "this paper".
\end{Verbatim}

\noindent Example Prompt 2 (\textit{after} adaption):
\begin{Verbatim}[breaklines,frame=single,numbers=left,xleftmargin=5mm]
You are a helpful assistant.
Title: EvalLM: Interactive Evaluation of Large Language Model Prompts on User-Defined Criteria
Abstract: {By simply composing prompts, developers can prototype novel generative applications...[remainder of the abstract omitted for readability]}
The first paragraph of your answer should explain and specify the relationship between EvalLM: Interactive Evaluation of Large Language Model Prompts on User-Defined Criteria and Herding AI Cats: Lessons from Designing a Chatbot by Prompting GPT-3. Be informative.
* In this first paragraph of your answer, you must explain and specify how EvalLM: Interactive Evaluation of Large Language Model Prompts on User-Defined Criteria is related to Herding AI Cats: Lessons from Designing a Chatbot by Prompting GPT-3 in one short sentence.
* In this first paragraph of your answer, you must start with "This paper might be related to Herding AI Cats: Lessons from Designing a Chatbot by Prompting GPT-3 because".
* In this first paragraph of your answer, you must specify how EvalLM: Interactive Evaluation of Large Language Model Prompts on User-Defined Criteria cites Herding AI Cats: Lessons from Designing a Chatbot by Prompting GPT-3 in one short sentence. The content from EvalLM: Interactive Evaluation of Large Language Model Prompts on User-Defined Criteria that cites Herding AI Cats: Lessons from Designing a Chatbot by Prompting GPT-3: {However, most designers (P1-7) mentioned how they were unsure about how they should revise their prompts' wellknown challenge with LLM prompts [72]., However, as the space of possible natural language instructions is near infinite, designers need to test as many possibilities as possible to identify high-performing prompts [38, 72].", "As LLMs are non-deterministic and even partial changes in a prompt can significantly influence generated outputs [37, 41], designers need to iterate on their prompts multiple times to achieve satisfactory results [27, 39, 57, 69, 72, 73]., to automatically generate, evaluate, and revise outputs that satisfy the criteria--without the designer needing to "herd" the LLM themselves [72].}.
* The first paragraph should have no more than 425 characters.
Your answer should replace EvalLM: Interactive Evaluation of Large Language Model Prompts on User-Defined Criteria with "this paper".
\end{Verbatim}

\noindent Output of example Prompt 2: \texttt{This paper might be related to Herding AI Cats: Lessons from Designing a Chatbot by Prompting GPT-3 because it cites the challenges in revising prompts, the need for extensive testing, and the iterative nature of achieving satisfactory results, referencing the concept of "herding" LLMs as discussed in the cited work.}

\vspace{1em}

\noindent Prompt 3 in Fig.~\ref{fig:prompts} (\textit{before} adaption):
\begin{Verbatim} [breaklines,frame=single,numbers=left,xleftmargin=5mm]
You are a helpful assistant in finding the relationships between two papers.
  Title: [TITLE_OF_RECOMMENDED_PAPER]
  Abstract:  [ABSTRACT_OF_RECOMMENDED_PAPER]
  Title: [TITLE_OF_PAPER_SHARED_BY_RELEVANT_USER]
  Abstract: [ABSTRACT_OF_RELEVANT_USER_SHARED_PAPER]
  Your answer should explain and specify how [TITLE_OF_RECOMMENDED_PAPER] is related to and different from [TITLE_OF_PAPER_SHARED_BY_RELEVANT_USER] with no more than 300 characters. If two papers are irrelevant, you should answer "NONE". Your answer must start with "This paper is related to '[TITLE_OF_PAPER_SHARED_BY_RELEVANT_USER]' because both papers". Be informative.
\end{Verbatim}

\noindent Example Prompt 3 (\textit{after} adaption):
\begin{Verbatim} [breaklines,frame=single,numbers=left,xleftmargin=5mm]
You are a helpful assistant in finding the relationships between two papers.
Title: EvalLM: Interactive Evaluation of Large Language Model Prompts on User-Defined Criteria
Abstract: {By simply composing prompts, developers can prototype novel generative applications...[remainder of the abstract omitted for readability]}
Title: Safer-Instruct: Aligning Language Models with Automated Preference Data
Abstract: {Reinforcement Learning from Human Feedback (RLHF) is a vital strategy for enhancing model...[remainder of the abstract omitted for readability]}
Your answer should explain and specify how EvalLM: Interactive Evaluation of Large Language Model Prompts on User-Defined Criteria is related to and different from Safer-Instruct: Aligning Language Models with Automated Preference Data with no more than 300 characters. If two papers are irrelevant, you should answer "NONE". Your answer must start with "This paper is related to Safer-Instruct: Aligning Language Models with Automated Preference Data because both". Be informative.
\end{Verbatim}

\noindent Output of example Prompt 3: \texttt{This paper is related to Safer-Instruct: Aligning Language Models with Automated Preference Data because both focus on improving LLMs' outputs. EvalLM refines prompts via user feedback, while Safer-Instruct automates preference data generation for safer AI.}

\vspace{1em}

\noindent Prompt 4 in Fig.~\ref{fig:prompts} (\textit{before} adaption):
\begin{Verbatim}[breaklines,frame=single,numbers=left,xleftmargin=5mm]
You are a helpful assistant for paper summarization.
[PROMPT_123_OUTPUTS]
First, you are required to shorten the above content with no more than [CHARACTER_LIMIT] characters. Note that:
- The shortened content must contain the following strings: [TITLE_OF_RELEVANT_PAPER]
- The shortened content must contain the following strings: [TITLE_OF_PAPER_SHARED_BY_RELEVANT_USER]
- The shortened content must contain the following strings: [RELEVANT_USER_SLACK_ID]
- Two papers are mentioned in [PROMPT_2_OUTPUT]. When you specify people's reactions or comments about [TITLE_OF_RELEVANT_PAPER], you should focus more on who reacted or commented and how these reactions or comments infer the potential reactions or comments about another paper based on the two papers' similarities than people's reactions or comments about [TITLE_OF_RELEVANT_PAPER].
- Do not remove any person's name (with or without '@'), institution's name, number, and conference/journal's name.
- Do not change the content's tone when it is low-confidence.
Second, you are required to bold at most three key phrases of the shortened content by adding ONE '*' to the left of the bolded text and ONE '*' to the right of the bolded text.
You should only bold the following text:
- The text tells what a paper is about.
- The text explains why a paper might be related to another paper or why a user might be interested in a paper, which is often after (does not include) the keywords such as "because" and "due to".
You should not bold the following text:
- [TITLE_OF_RELEVANT_PAPER]
- [TITLE_OF_PAPER_SHARED_BY_RELEVANT_USER]
- [RELEVANT_USER_SLACK_ID]
\end{Verbatim}

\noindent Example Prompt 4 (\textit{after} adaption):
\begin{Verbatim} [breaklines,frame=single,numbers=left,xleftmargin=5mm]
You are a helpful assistant for paper summarization.
{Congratulations to @tsk and team, including Juho Kim, on developing EvalLM! This system aids in refining LLM prompts by evaluating outputs against user-defined criteria, showing a 59% reduction in needed revisions. It enables examining twice as many outputs, enhancing prompt development efficiency.  This paper might be interesting to @xinyiz because @xinyiz've liked Safer-Instruct: Aligning Language Models with Automated Preference Data; improving LLMs' outputs. EvalLM refines prompts via user feedback, while Safer-Instruct automates preference data generation for safer AI.}
First, you are required to shorten the above content with no more than 386 characters. Note that:
- The shortened content must contain the following strings: Safer-Instruct: Aligning Language Models with Automated Preference Data
- The shortened content must contain the following strings: @anonymity
- Do not remove any person's name (with or without '@'), institution's name, number, and conference/journal's name.
- Do not change the content's tone when it is low-confidence.
Second, you are required to bold at most three key phrases of the shortened content by adding ONE '*' to the left of the bolded text and ONE '*' to the right of the bolded text.
You should only bold the following text:
- The text tells what a paper is about.
- The text explains why a paper might be related to another paper or why a user might be interested in a paper, which is often after (does not include) the keywords such as "because" and "due to".
You should not bold the following text:
- Safer-Instruct: Aligning Language Models with Automated Preference Data
- @anonymity
\end{Verbatim}

\noindent Output of example Prompt 4, where the title in blue is clickable and links to the article's Semantic Scholar page in our deployment: \texttt{Congrats to @tsk, Juho Kim on \textbf{EvalLM}, reducing prompt revisions by 59\% and doubling output examination. It refines LLM prompts with user feedback. This might interest @anonymity, who liked \textcolor{blue}{EvalLM: Interactive Evaluation of Large Language Model Prompts...}, for its focus on \textbf{improving LLM outputs} through automated preference data for safer AI.}

\section{Social signals used in PaperPing}
See Table~\ref{tab:paperping_heuristics}.
\begin{table*}[t]
\centering
\scriptsize
\caption{Social signals used in PaperPing}
\label{tab:paperping_heuristics}
\resizebox{\textwidth}{!}{%
\begin{tabular}{|p{2cm}|p{3.75cm}|p{3.75cm}|}
\hline
\textbf{Category} & \textbf{Heuristic} & \textbf{Example message} \\ \hline
\multirow{4}{2cm}{\textbf{Recommended Paper Information (I)} \textit{This paper is about...}} 
& \textbf{h1 -} Is the paper's author a channel member? \textit{(i.e., Does the author show up at least once in the past 3 months among papers posted, reacted to, or replied to by at least one human user?)} 
& \textit{Congrats @\{UserSlackID\}} \\ \cline{2-3} 
& \textbf{h2 -} Was the paper's author mentioned recently in the channel? 
& \textit{\{AuthorName\} also authored \{K\} papers that you've discussed in the past three months in the channel} \\ \cline{2-3} 
& \textbf{h3 -} Does the author's affiliation overlap with a channel member's affiliation? 
& \textit{A paper from \{AuthorAffiliation\}} \\ \cline{2-3} 
& \textbf{h4 -} Was the paper's venue mentioned recently in the channel? \textit{(i.e., Does the venue show up at least once in the past 3 months among papers posted, reacted to, or replied to by at least one human user?)} 
& \textit{You've reacted positively to other papers from CHI} \\ \hline

\multirow{3}{2cm}{\textbf{Paper Connections (II)} \textit{This paper might be related to another paper in the channel, because...}} 
& \textbf{h5 -} What is the relationship between two papers? \textit{(i.e., Does one cite the other? Do they have shared authors? Are two papers semantically relevant?)} 
& \textit{Cites: The Participatory Turn in AI Design: Theoretical Foundations and the Current State of Practice (@anonymity shared in a thread), for its methodology: This shift in design strategy transformed the participatory design’s user-centered goal from "consulting" to "ownership," aligning with the principles outlined in {[}10, 25{]}} \\ \cline{2-3} 
& \textbf{h6 -} How did channel members react and reply to another paper? \textit{(If two papers are relevant, such reactions and replies could infer how channel members would react and reply to the recommended paper.)} 
& \textit{Shared authors: 3 paper authors, Tae Soo Kim, Juho Kim, and Yoonjoo Lee, also contributed to Cells, Generators, and Lenses: Design Framework for Object Oriented Interaction with Large Language Models that @anonymity1 shared in a thread, with 1 reply from @anonymity2 ("I like that the paper...")} \\ \cline{2-3} 
& \textbf{h7 -} Was another paper authored by a channel member? 
& \textit{Related paper: Redefining Qualitative Analysis in the AI Era: Utilizing ChatGPT for Efficient Thematic Analysis that @scholar shared in a thread which received 2 replies and 3 reactions} \\ \hline

\multirow{2}{2cm}{\textbf{User Connections (III)} \textit{This paper might be interesting to someone in the channel, because...}} 
& \textbf{h8 -} Is the paper relevant to a channel member? \textit{(i.e., Is the paper's semantic embedding similar to that of papers already of interest to a channel member?)} 
& \textit{Possibly of interest to @anonymity} \\ \cline{2-3} 
& \textbf{h9 -} What is the relationship between a user's interests and the recommended paper? \textit{(i.e., Has the user liked a previous paper that's similar? Cites or is cited by this paper? Shares the same authors?)} 
& \begin{tabular}[t]{@{}p{3.75cm}@{}} 
\textit{Possibly of interest to @anonymity, because...} \\ 
- \textit{you've liked several similar papers in the channel} \\ 
- \textit{you've liked several of James's papers in the channel} \\ 
- \textit{you've liked several CHI papers in the channel} \\ 
- \textit{several of your publications are similar} \\ 
- \textit{you've cited similar papers before} 
\end{tabular} \\ \hline
\end{tabular}
}
\end{table*}

\section{Numbers of Interactions in Each Channel}

See Table~\ref{tab:log_data_results}.

\begin{table*}[]
\centering 
\caption{Numbers of Interactions in Each Channel}
\label{tab:log_data_results}
\begin{tabular}{|r|r|r|r|r|r|r|r|}
\toprule
Channel index & \begin{tabular}[c]{@{}l@{}}\# human \\ recs before \\ PaperPing\end{tabular} & \begin{tabular}[c]{@{}l@{}}\# human \\ recs after \\ PaperPing\end{tabular} & \begin{tabular}[c]{@{}l@{}}\#  emojis\\ before \\ PaperPing\end{tabular} & \begin{tabular}[c]{@{}l@{}}\# emojis\\ after \\ PaperPing\end{tabular} & \begin{tabular}[c]{@{}l@{}}\#  comments\\ before \\PaperPing\end{tabular} & \begin{tabular}[c]{@{}l@{}}\# comments\\ after \\ PaperPing\end{tabular} & \begin{tabular}[c]{@{}l@{}}\# \\ PaperPing \\ recs\end{tabular} \\ \midrule
1 & 31 & 6 & 27 & 21 & 12 & 4 & 51 \\ \hline
2 & 18 & 4 & 0 & 8 & 0 & 2 & 65 \\ \hline
3 & 2 & 7 & 4 & 27 & 5 & 4 & 20 \\ \hline
4 & 3 & 2 & 0 & 1 & 0 & 0 & 12 \\ \hline
5 & 4 & 2 & 3 & 5 & 0 & 0 & 69 \\ \hline
6 & 26 & 25 & 22 & 20 & 2 & 0 & 67 \\ \hline
7 & 20 & 89 & 23 & 127 & 2 & 14 & 64 \\ \hline
8 & 19 & 10 & 18 & 14 & 6 & 4 & 54 \\ \hline
9 & 3 & 1 & 0 & 3 & 0 & 0 & 102 \\ \hline
10 & 2 & 8 & 0 & 21 & 0 & 7 & 106 \\ \hline
11 & 2 & 0 & 0 & 0 & 0 & 1 & 15 \\ \hline
12 & 7 & 18 & 5 & 12 & 4 & 2 & 62 \\ \hline
14 & 6 & 13 & 2 & 13 & 0 & 0 & 71 \\ \hline
15 & 233 & 41 & 249 & 40 & 183 & 41 & 6 \\ \hline
16 & 7 & 0 & 4 & 10 & 0 & 0 & 22 \\ \hline
17 & 19 & 0 & 34 & 3 & 3 & 0 & 21 \\ \hline
18 & 12 & 0 & 22 & 0 & 11 & 3 & 25 \\ \bottomrule

\end{tabular}
\end{table*}

\end{document}